\documentclass[11pt]{article}  
\usepackage{graphicx}
\newcommand{\be}{\begin{equation}}
\newcommand{\ee}{\end{equation}}
\newcommand{\bea}{\begin{eqnarray}}
\newcommand{\eea}{\end{eqnarray}}

\begin{document}
\vspace{0.5in}
\vspace{.5in} 
\begin{center} 
{\LARGE{\bf Kink Solutions With Power Law Tails}} 
\end{center} 

\vspace{0.2in} 

\begin{center}
{{\bf Avinash Khare}} \\
{Physics Department, Savitribai Phule Pune University \\
 Pune 411007, India}
\end{center}

\begin{center}
{{\bf Avadh Saxena}} \\ 
{Theoretical Division and Center for Nonlinear Studies, 
Los Alamos National Laboratory, Los Alamos, New Mexico 87545, USA}
\end{center}

\vspace{0.1in}

{\bf Table of Contents}

1. Introduction

2. Kink formalism

3. Kinks with power law tails of the form $eppe$

4. Kinks with power law tails of the form $peep$

5. Kinks with power law tails of the form $pppp$

6. Kinks with power-tower tails

7. Kinks with kink tails of the form $eeep$

8. Kinks with kink tails of the form $pppe$

9. Explicit kink solutions with power law tails

10. Kink-kink and kink-antikink forces for power law tails 

11. Kink-antikink collisions at finite velocity 

12. Open problems

\noindent{\bf {Abstract:}}
We present a comprehensive review about the various facets of kink solutions
with a power law tail which have received considerable attention during the last
few years. This area of research is in its early stages and while several
aspects have become clear by now, there are a number of issues which have 
only been partially understood or not understood at all. We first discuss the
aspects which are reasonably well known and then address in some detail the 
issues which are only partially or not understood at all. We present a wide
class of higher (than sixth) order field theory models admitting implicit kink 
as well as mirror kink solutions where the two tails facing each other have a 
power law or a power-tower type fall off while the other two ends not facing 
each other could have either an exponential or a power law tail. The models 
admitting implicit kink solutions where the two ends facing each other have an  
exponential tail while the other two ends have a power law tail are also 
discussed. Moreover, we present several field theory models which admit 
explicit kink solutions with a power law fall off. We note that in all the 
polynomial models while the potential $V(\phi)$ is continuous, its derivative is 
discontinuous. We also discuss one of the most important and only partially 
understood issues of the kink-kink and the kink-antikink forces in case the 
tails facing each other have a power law fall off. Finally, we briefly discuss 
the kink-antikink collisions at finite velocity and present some open questions.

\section{Introduction} 

Recently it was found that certain $1+1$ dimensional higher order field 
theories, including $\phi^8, \phi^{10}$ and $\phi^{12}$ models, admit kink 
solutions with a power law tail at either both the ends or a power law tail at 
one end and an exponential tail at the other end \cite{KCS, Chapter, bmm}. 
An example of the latter is the octic potential first studied in the context of 
massless mesons \cite{Lohe79} as well as some other studies related to the 
kink solutions with a long range tail \cite{gonz1, gonz2, Mello, Guerrero, 
GLG, baz06, Gomes, gani1}. This is in contrast to 
almost all the kink 
solutions that have been discussed during the last four decades where the kink 
solutions have an exponential tail at both the ends 
\cite{raj, manton2, vil, vach, shnir}, 
the prototype being the celebrated $\phi^4$ kink.  We provide, however, 
an example  of a $\phi^6$ kink with a power law tail \cite{Gomes} in Sec. 9. The study of higher order field 
theories, their attendant kink excitations as well as the associated kink interactions 
and scattering are important in a variety of physical contexts ranging from 
successive phase transitions \cite{KCS, Chapter, Gufan82, Gufan78} to 
isostructural phase transitions \cite{Pavlov99} to models involving 
long-range interaction between massless mesons \cite{Lohe79}, as well as from 
protein crystallization \cite{Boulbitch97} to successive phase transitions 
presumably driving the late time expansion of the universe \cite{Greenwood09}. 
Thus, understanding kink behavior in these models provides useful insights into 
the properties of domain walls in materials, condensed matter, high energy 
physics, biology and cosmology.

The discovery of these power law kinks has raised several interesting questions 
such as the strength and the range of the kink-kink (KK) and the kink-antikink 
(K-AK) forces \cite{Manton19, cddgkks}, the possibility of resonances \cite{christov} 
and scattering \cite{gani2, gani21, chr21, camp}, stability analysis of such kinks 
\cite{Gomes}, explicit kink solutions with power law tail, etc. From this perspective, 
it is worth noting that the celebrated Manton's method \cite{manton2, manton1} 
using 
collective coordinates provides the answer for both the strength and the range of the 
KK and K-AK interactions in case they have an exponential tail facing each other. 
By now several aspects of the kink solutions with a power law tail have been 
understood while many issues are either only partially or not understood at all.
We reckon that it is now the appropriate time to provide a comprehensive review 
of those aspects which are reasonably well understood and also clearly bring 
out the issues which are either only partially or not understood at all and 
deserve further attention. 

Some of the key issues related to the kink solutions with power law tails are 
as follows. 

\begin{enumerate}

\item	What are the signatures of the kink solutions with a power
law tail in contrast to the kink solutions with an exponential tail? 

\item What are the various possible types of models where the kink tail at 
either one end or both the ends has either a power law or an effective power 
law fall off?

\item Are there explicit kink solutions with a power law tail at either one or
both the ends?

\item Can one estimate the kink-kink and the kink-antikink forces in case 
these kink solutions have power law tails? How do these forces compare to the
corresponding KK and K-AK forces in case the kink solutions
have exponential tails? Besides, what is the ratio of the magnitude of the 
K-AK and KK forces in such cases? Note that for the 
exponential tail case, this ratio is one.
		
\item Is there a bound state formation and are there escape windows when one 
considers the collision of  the kink and the antikink
with power law tails at finite velocity and if yes, is it universal? 

\end{enumerate}
The purpose of this review article is to provide answers to some of the 
questions raised above and clearly spell out the issues which are either only 
partially understood or not understood at all so far. 

The plan of the review is the following. In Sec. II we first set up the 
notations and show that there is always an underlying supersymmetry in the
problem when we set up the Schr\"odinger-like kink stability equation. We also 
show that sometimes it is more convenient to  consider the 
Schr\"odinger-like stability equation
in the field variable $\phi$ rather than the coordinate $x$.
We then give a recipe for constructing kink solutions with a power law tail. 
Next, we show that in the case of the kink solutions with a power law tail at 
either one or both the
ends, there is no gap between the zero mode and the beginning of the continuum
in the Schr\"odinger-like stability equation. In addition, we consider the 
case of two adjoining kinks and point out the various possible forms for the
kink tails in the two adjoining kink case. 

In Sections III to IX we consider
distinct possible one-parameter family of potentials corresponding to the 
various possible forms for the two adjoining kinks with at least one power law
tail. In particular, in Sections III to VI we discuss one-parameter family of
potentials admitting a kink from $0$ to $1$ and a mirror kink from $-1$ to $0$ 
and the corresponding two antikinks where either two or all four of the kink tails 
have a power law fall off. In Sec. III \cite{KS19} we present a one-parameter 
family of potentials of the form $\phi^{2n+2} (1 -\phi^2)^2$, $n = 1, 2, 3,...$ \,. 
For all the kink solutions of this model, while around $\phi = 0$ one has a power law tail, 
around $\phi = \pm 1$ one has an exponential tail.
In Sec. IV \cite{KS19} we present a one-parameter family of potentials of the
form $\phi^{2}(1 -\phi^2)^{2n+2}$,~ $n = 1, 2, 3,...$ \,.
For all the kink solutions of this model, while around 
$\phi = 0$ one has an exponential tail, around $\phi = \pm 1$ one has a 
power law tail. In Sec. V \cite{KS19} we present a one-parameter family of 
potentials of the form $\phi^{2m+2}(1 -\phi^2)^{2n+2}$,~ $n,m \ge 1$, the 
kink solutions of which have a power law tail 
around both $\phi = 0$ as well as $\phi = \pm 1$. 
In Sec. VI \cite{KS20} we present a one-parameter family of potentials of the 
form $\phi^{2m+2} [(1/2)\ln(\phi^2)]^2$~$, m = 1, 2, 3,...$ \,, for the kink solution of 
which while 
around $\phi = 0$ one has a power-tower tail (which is effectively a power law 
type tail), around $\phi = \pm 1$ one has an exponential tail. 
Further, in the same section we also present a two-parameter family of potentials
of the form $\phi^{2m+2} [(1/2)\ln(\phi^2)]^{2n+2}$~$, m, n = 1, 2, 3,...$\,, 
for the kink solution of which while 
around $\phi = 0$ one has a power-tower tail (which is effectively a power law 
type tail), around $\phi = \pm 1$ one has a power law tail. 

In Sections VII and VIII we present a one-parameter family of potentials 
admitting non-mirror kinks and the corresponding antikinks with kink tails of 
the form $pppe$ and $eeep$, respectively where $p$ and $e$ correspond to 
power law and exponential tail, respectively. In particular, 
in Sec. VII we present a one-parameter family of potentials of the form 
$V(\phi) = \frac{1}{2}(1-\phi^2)^{2n+2} (2-\phi^2)^{2}\,,~~n = 1, 2, 3,...$ 
which admit a kink solution from $-1$ to $1$ with a power law tail around both 
$\phi = -1$ and $\phi = 1$ and another kink solution from $1$ to $\sqrt{2}$ 
(and the corresponding mirror kink solution from $-\sqrt{2}$ to $-1$) with a 
power law kink tail around $\phi = 1$ but an exponential tail around 
$\phi = \sqrt{2}$. 
In Sec. VIII we present a one-parameter family of potentials of the form 
$V(\phi) = \frac{1}{2}(1-\phi^2)^{2} (2-\phi^2)^{2n+2}\,,~~n = 1, 2, 3,...$ 
which admit a kink solution from $-1$ to $1$ with an exponential tail around both
$\phi = -1$ and $\phi = 1$ and another kink solution from $1$ to $\sqrt{2}$ 
(and the corresponding mirror kink from $-\sqrt{2}$ to $-1$) with an 
exponential kink tail around $\phi = 1$ but a power law tail around 
$\phi = \sqrt{2}$. 

Unfortunately, all the kink solutions discussed in Sections III to VIII are 
only in an implicit form. In Sec. IX we discuss three different models 
for which explicit kink solutions with a power law tail can be obtained. 
In particular, we discuss two different one-parameter 
family of potentials of the form $\phi^{2n+2}|1-\phi^{2n}|^{3}$ and 
$|1-\phi^{2n}|^{(2n+1)/n}$, where $n = 1, 2, 3,...$ for which explicit kink 
solutions with power law tails can be obtained \cite{KS21}. We note that in 
both these models, while the potential $V(\phi)$ is continuous around 
$\phi = \pm 1$, its derivative is not continuous. We also discuss one 
nonpolynomial model in which an explicit kink solution with a power law tail
can be obtained \cite{mra,ko}. The nice thing about this model is that in this
model the potential $V(\phi)$ and its derivative are both continuous. 
In Sec. X we discuss what we consider to be the most 
important (and not so well understood) issue of the KK and 
K-AK forces in the case of kink solutions with power law tails.
Following the seminal paper of Manton \cite{Manton19} we show that both the KK 
and K-AK forces have a power law fall off in 
contrast to the exponentially small KK and K-AK forces in 
the case of the exponential tail. Further, it turns out that while the ratio
of the magnitude of the K-AK force to the KK force is always one in
the case of the models with exponential tails, this ratio is in fact always
less than one and progressively decreases as the kink tail becomes 
progressively longer \cite{cddgkks}. In Sec. XI  we discuss the question of the 
kink-antikink collisions at finite (nonzero) velocity in the case 
of the kinks with a power law tail. Finally, in Sec. XII we highlight some
of the major issues which are either only partially or not understood at all.

\section{Formalism}

Consider a relativistic, neutral scalar field theory in $1+1$ dimensions with
the Lagrangian density
\be\label{2.1}
{\mathcal L} = \frac{1}{2}\left(\frac{\partial \phi}{\partial t}\right)^2  
- \frac{1}{2}\left(\frac{\partial \phi}{\partial x}\right)^2 - V(\phi)\,,
\ee
which leads to the equation of motion
\be\label{2.2}
\left(\frac{\partial^2 \phi}{\partial t^2}\right)  
- \left(\frac{\partial^2 \phi}{\partial x^2}\right) = - \frac{dV}{d\phi}\,.
\ee
We are working in the Minkowski space and will use the metric $\eta_{\mu \nu}
= diag(+1, -1)$. Further, since the Lagrangian does not explicitly depend on 
the spacetime coordinates, by invoking Noether's theorem it follows that 
there is a conserved energy-momentum tensor,
\bea\label{2.2a}
&&T_{\mu \nu} = \frac{\partial {\mathcal {L}}}{\partial(\partial^{\mu} \phi)}
\partial_{\nu} \phi - \eta_{\mu \nu} {\mathcal {L}} \nonumber \\
&& = \partial_{\mu} \phi \partial_{\nu} \phi - \eta_{\mu \nu} {\mathcal {L}}
\,,
\eea
with $\partial^{\mu} T_{\mu \nu} = 0$. Thus, the energy density $E$ and the 
momentum density $P$ can be immediately obtained from the components of the
energy momentum tensor, i.e. 
\be\label{2.2b}
T_{00} = E = \frac{1}{2}(\dot{\phi})^2 + \frac{1}{2}(\phi')^{2} +V(\phi)\,, 
\ee
\be\label{2.2c}
T_{01} = \dot{\phi} \phi'  = -P\,.
\ee
Here $\dot{\phi}$ and $\phi'$ correspond respectively to the time and the space
derivative of $\phi$. We assume that the potential $V(\phi)$ is smooth and 
non-negative. Thus 
$V(\phi)$ attains its global minimum value of $V(\phi) = 0$ for one or more 
values of $\phi$ which are the global minima of the theory. We shall choose 
$V(\phi)$ such that it has two or more global minima so that one has static 
kink and antikink solutions interpolating between the two adjacent global 
minima as $x$ increases from $-\infty$ to $+\infty$. 

Since the neutral relativistic scalar field theory as given by Eq. (1) is
Lorentz invariant, once the static kink solution is known, the corresponding
time-dependent solution is easily obtained by Lorentz transformation. Hence
it is enough to look for the static kink solution of the field equation
\be\label{x}
\left(\frac{d^2 \phi}{d x^2}\right) = \frac{dV}{d\phi}\,.
\ee
On integrating Eq. (\ref{x}) and using the fact that for the kink solution
$V(\phi)$ as well as $\frac{d\phi}{dx}$ vanish at the degenerate global
minima, say $\phi = a$ and $\phi = b$, one obtains the first order ODE
\be\label{2.3}
\frac{d\phi}{dx} = \pm \sqrt{2V(\phi)}\,.
\ee
This is a special case of the Bogomolnyi technique \cite{bog} (although known
much earlier in this context). Here the equations with $+$ and $-$ sign 
are called the self-dual and anti-self-dual field equations, respectively. We 
shall refer to the first order equation as the Bogomolnyi equation for the kink. 

The corresponding static kink energy $E$ (which also equals the corresponding 
antikink energy) and which is also referred to as the kink mass $M_K$ is given
by
\be\label{2.4}
M_K = \int_{-\infty}^{\infty} \bigg (\frac{1}{2} 
\left(\frac{d\phi}{dx}\right)^2  +V(\phi) \bigg )\, dx\,.
\ee
In view of the first order Eq. (\ref{2.3}), the kink mass $M_K$ takes a 
simpler form
\be\label{2.5}
M_K = \int_{\phi=a}^{\phi=b} {2V(\phi)}\, d\phi\,,
\ee
where as $x$ goes from  $-\infty$ to +$\infty$, the kink solution goes from
one minimum at $\phi= a$ to the adjacent minimum at $\phi= b$. 
Since we are considering a relativistic  neutral scalar field theory, once a
static kink solution is known, the corresponding moving kink solution is
immediately obtained by a Lorentz boost. 

The recipe for constructing kink solutions with a power law tail or an exponential 
tail is clear and well known. Since a kink solution has finite energy it implies that 
the solution must approach one of the minima (vacua) $\phi_0$ of the theory as
$x$ approaches either +$\infty$ or $-\infty$. If the lowest non-vanishing derivative 
of the potential at the minimum has order $m$, then by Taylor expanding the 
potential at the minimum and writing the field close to it as $\phi = \phi_0 + \eta$, 
one finds that the self-dual (or Bogomolnyi) first order equation in $\eta$ implies 
that (assuming that the potential vanishes at the minimum)
\be\label{2.5g}
\frac{d\eta}{dx} \propto \eta^{m/2}\,.
\ee
Thus if $m =2$ then $\eta \propto e^{-\alpha x}$ so that the kink tail has
an exponential fall off while if $m > 2$ then 
$\eta \propto 1/x^{2/(m-2)}$ so that it is a power law kink tail. 
This recipe has been used to construct 
several one-parameter family of potentials with various possible forms of
the power law and the exponential tails.    

As is well known, kink is a topological object. In particular, there is an underlying 
current which is conserved by construction while the corresponding topological 
charge is nonzero in the case of kink solutions. In particular, the corresponding 
conserved current is
\be\label{2.5a}
J_{\mu}(x) = \epsilon_{\mu \nu} \frac{d\phi}{dx_{\nu}}\,,~~~\mu, \nu = 0, 1\,.
\ee
Hence for the kink solution, the topological charge density is 
\be\label{2.5b}
J_{0}(x) = \frac{d\phi}{dx} = \sqrt{2V(\phi)}\,,
\ee
where Eq. (\ref{2.3}) has been used in writing the second equality. 
Thus, the topological charge $Q$ is given by 
\be\label{2.5c}
Q = \int_{a}^{b} d\phi = b-a\,.
\ee

For the kink solution one can perform the linear stability analysis by 
considering
\be\label{2.6}
\phi(x,t) = \phi_k(x) + \psi(x) e^{i\omega t}\,,
\ee
where $\phi_k$ is the kink solution. 
On substituting $\phi(x,t)$ as given by Eq. (\ref{2.6}) in the
field Eq. (\ref{2.2}) and retaining terms of order $\psi$, it is easily shown 
that $\psi(x)$ satisfies a Schr\"odinger-like equation
\be\label{2.7} 
-\frac{d^2 \psi(x)}{dx^2} +V(x)\psi(x) = \omega^2 \psi(x)\,,
\ee
where
\be\label{2.7a}
V(x) = \frac{d^2 V(\phi)}{d\phi^2}\Big|_{\phi = \phi_{k}(x)}\,.
\ee
Here $\phi_{k}(x)$ denotes the corresponding kink (or antikink) solution. It
is well known that the stability Eq. (\ref{2.7}) always admits a zero-mode, i.e.
\be\label{2.8}
\omega_0 = 0\,,~~\psi_0 (x) = \frac{d\phi_k(x)}{dx}\,,
\ee
where because of the Bogomolnyi equation (\ref{2.3}) it is clear that 
$\psi_0(x)$ is indeed nodeless, thereby guaranteeing the linear stability of 
the kink solution of any theory. 

\subsection{Underlying Supersymmetry in the Stability Equation}

Let us now show that there is an underlying supersymmetry in the kink 
stability equation (16). This is because, as is well known from the supersymmetric 
quantum mechanics formalism \cite{CKS}, for the Schr\"odinger-like Eq. (\ref{2.7})  
if the corresponding ground state eigenfunction $\psi_0(x)$ is nodeless 
then there is always an underlying supersymmetry in the problem which 
is unbroken. In particular, in that case the corresponding superpotential 
$W(x)$ is given by
\be\label{2.9}
W(x) = -\frac{\psi_{0}'(x)}{\psi_{0}(x)}\,.
\ee
Further, in terms of the superpotential $W(x)$, the corresponding potential
$V_-(x) \equiv V(x)$ is given by
\be\label{2.10}
V(x) \equiv V_{-}(x) = W^2(x) - W'(x)\,,
\ee
while the corresponding partner potential $V_{+}(x)$ with one less bound state 
compared to $V_{-}(x)$ is given by
\be\label{2.11}
V_{+}(x) = W^2(x) + W'(x)\,.
\ee
Using Eqs. (\ref{2.3}) and (\ref{2.8}) in Eq. (\ref{2.9}) we can rewrite 
$W(x)$ as
\be\label{2.12}
W(x) = -\frac{V'(\phi)}{\sqrt{2V(\phi)}}\Big|_{\phi = \phi_{k}(x)}\,, 
\ee
so that as expected $V(x)$ (i.e. $V_{-}(x)$) is as given by Eq. (\ref{2.7}) 
while the corresponding partner potential $V_{+}(x)$ with one less bound state 
is given by
\be\label{2.13}
V_{+}(x) = \bigg (\frac{[V'(\phi)]^2}{V(\phi)} 
-V''(\phi) \bigg )\Big|_{\phi = \phi_{k}(x)}\,.
\ee
This is interesting because while doing the stability analysis for a kink 
solution, if one obtains more modes (called the breather modes) than just the 
zero mode (also called the translation mode), then using the supersymmetry 
(SUSY) formalism one can obtain another kink potential with at least a zero 
mode. As an illustration, for the famous double-well $\phi^4$ potential it is well 
known that if one does the stability analysis, then one has a breathing mode 
apart from the translation mode. By following this formalism, it is straightforward 
to discern that the corresponding supersymmetric kink potential $V_{+}$ with 
only the translation mode in the stability analysis is the celebrated sine-Gordon 
model thereby showing a remarkable connection between the two distinct kink 
bearing models, $\phi^4$ and the sine-Gordon field theory, the former is a 
non-integrable and the latter is an integrable model. On the other hand, since
for kink solutions with a power law tail at one or both the ends, there is
only the zero mode and no other vibrational mode, it follows that the partner
potential $V_{+}$ as given by Eq. (22) in such cases will not admit a kink 
solution. 

\subsection{Stability Equation in Terms of the Field $\phi$}

We now show that the kink stability equation as given by Eq. (\ref{2.7}) 
which is cast in terms of the eigenfunctions $\psi(x)$ can
also be recast in terms of $\psi(\phi)$. To that end, we start from 
Eq. (\ref{2.7}) and using the Bogomolnyi Eq. (\ref{2.3}) we obtain
\be\label{2.14}
\psi'(x) = \psi'(\phi) \frac{d\phi}{dx} = \pm \sqrt{2V(\phi)} \psi'(\phi)\,.
\ee
Further
\be\label{2.15}
\psi''(x) = 2V(\phi) \psi''(\phi) +V'(\phi)\psi'(\phi)\,.
\ee
On using Eqs. (\ref{2.14}) and (\ref{2.15}) in the stability Eq. (\ref{2.7}), 
we find that in terms of $\psi(\phi)$ the stability equation takes the form
\be\label{2.16}
-2V(\phi)\psi''(\phi) - V'(\phi) \psi'(\phi) + V''(\phi) \psi(\phi) 
= \omega^2 \psi(\phi)\,.
\ee
It is straightforward to verify that for any kink bearing potential $V(\phi)$
there is always a zero mode which is given by
\be\label{2.17}
\psi_0(\phi) \propto \sqrt{V(\phi)}\,.
\ee

The obvious question is what is the advantage of casting the stability 
equation in terms of $\psi(\phi)$ rather than in terms of $\psi(x)$? One 
advantage is in the context of those cases where the kink solution is only 
implicitly but not explicitly known. As an illustration, almost all the kink solutions 
with a power law tail are only implicitly known. In such cases we do not know the 
explicit form of the zero
mode $\psi(x)$. However, in view of Eq. (\ref{2.17}) one always knows the form
of the zero mode $\psi_0(\phi)$. But what is even more interesting, sometimes 
it so happens that in case there are breathing modes in addition to the 
translational zero mode in the stability equation, at times 
it is easier to guess the form of the excited state eigenfunction 
$\psi_{n}(\phi),~ n > 0$ instead of the form of the corresponding eigenfunction 
$\psi_n(x)$. One such famous example is the second 
excited state of the stability equation in the case of the kink solution for 
the $\phi^6$ field theory \cite{clee} characterized by 
\be\label{2.18}
V(\phi) = (\phi^2+\epsilon^2)(1-\phi^2)^2\,.
\ee
As was pointed out a long time ago by Christ and Lee \cite{clee}, in the kink 
stability equation for this case, if $\epsilon^2 = 1/2$, the second 
excited state eigenfunction and the corresponding eigenvalue are
\be\label{2.19}
\psi_2(\phi) = \sqrt{1-\phi^2}(\phi^2 -1/4)\,,~~~\omega_2 = \sqrt{3}/2\,.
\ee

\subsection{No Gap Between Zero Mode and the Beginning of Continuum for Kink 
Solutions with a Power Law Tail}

We now show that if there is a kink solution with a power law tail at either both
the ends or at one of the two ends, then there is no gap between the zero mode 
and the beginning of the continuum in the corresponding Schr\"odinger-like 
stability Eq. (\ref{2.7}). The proof is rather straightforward. 
Let us first discuss the case when there is a kink solution from $\phi = 0$
to $\phi = a$ as $x$ goes from $-\infty$ to +$\infty$, respectively with there 
being a power law tail around $\phi = 0$, i.e.
\be\label{2.20}
\lim_{x \rightarrow -\infty}  \phi(x) \simeq 0 + c x^{-\beta}\,,~~\beta > 0\,.
\ee
In view of Eq. (\ref{2.5g}) this implies that if there is a kink solution from 
$0$ to $a$ with a power law tail around $\phi = 0$, then around $\phi = 0$
the potential $V(\phi)$ must behave as
\be\label{2.21}
\lim_{\phi \rightarrow 0}  V(\phi) \simeq \phi^{(2+2\beta)/\beta}\,.
\ee
Using the fact that the potential $V(x)$ which appears in the stability
analysis of a kink solution of Eq. (\ref{2.7}) is given by Eq. (\ref{2.7a}),
and further using Eqs. (\ref{2.20}) and (\ref{2.21}) it then follows that as 
$x \rightarrow -\infty$, the potential $V(x)$ around $x \rightarrow -\infty$ is given by
\be\label{2.22}
V(x\rightarrow -\infty) \propto \lim_{x \rightarrow -\infty} \phi_{k}^{2/\beta}
= 0\,,
\ee
so that the continuum in the Schr\"odinger-like Eq. (\ref{2.7}) begins from
$\omega^2 = 0$, i.e. there is no gap between the zero mode and the beginning 
of the continuum \cite{LL}. The argument trivially goes through in case the
kink solution is from $\phi = a$ to $\phi = b$ as $x \rightarrow -\infty$ to
$x \rightarrow \infty$, respectively with power law tail around $\phi = a$ 
or/and $\phi = b$. This is because on expanding $V(\phi)$ around $\phi = a$ 
(or $b$ as the case may be) leads us to an equation essentially identical to 
Eq. (\ref{2.20}) and the argument again goes through.

It is worth pointing out that if instead one considers the stability equation in the 
case of either the exponential or the super-exponential kink tails \cite{PKS}, 
there is always a gap between the zero mode and the beginning of the
continuum. In fact, in these cases, depending on the model one can even have 
one or more extra bound states, called vibration modes, and there is always
a gap between the last vibrational mode and the beginning of the
continuum. This is to be contrasted with the kink solution with power law
tails for which the only discrete mode is the zero mode and there is no gap
between the zero mode and the beginning of the continuum. Such a zero
energy bound state is called a half bound state.   

\subsection{Various Possible Tails Between the Two Kink Solutions}

Using Eq. (\ref{2.5g}), the recipe for constructing models which can give kink
solutions with either a power law tail or an exponential tail is clear. In  
particular using this recipe several potentials have been 
constructed which admit a kink solution with a power law tail at both ends.
A typical example is the potential $V(\phi) = (1-\phi^2)^{2n+2}$ with $n = 1,
2, 3,...$ \cite{KS19}. Further, several one-parameter family of potentials have 
been constructed with kink and mirror kink solutions with various possible 
options for the kink tails. Let us denote the two adjoining kink solutions as 
kink $1$ and kink $2$ and without any loss of generality we will assume that 
kink $1$ is to the left of kink $2$. One has two kink tails corresponding to 
the kink solution $1$ (which we denote by $K_{1L}$ and $K_{1R}$) and two kink 
tails $K_{2L}$ and $K_{2R}$ for the kink solution $2$. In Table 1 we give all 
eight possible forms of the kink tails. The well studied case for almost five 
decades is when all the four tails (i.e. two tails of the first kink and the 
two tails of the second kink) have exponential fall off which we denote by  
$K_{1Le}, K_{1Re}, K_{2Le}, K_{2Re}$, respectively and for simplicity we will 
denote such tails simply as $eeee$. On the other hand, the recent study by 
several groups \cite{gani1, Manton19, christov, cddgkks} have concentrated on 
the case when the kink tails have the form $K_{1Le}, K_{1Rp}, K_{2Lp}, 
K_{2Re}$ and  we will denote this possibility as $eppe$. In fact, there are two 
other possibilities for which the kinks and the 
corresponding mirror kinks are also possible and these are of the form 
$peep$ and $pppp$. For the other four possibilities shown in Table 1, one 
necessarily has to consider non-mirror kinks since for them the kink tails are 
of the form $eeep$ and $peee$ (and without loss of generality one can consider 
one of these two possibilities, say $eeep$) and $pppe$ and $eppp$ (and again 
without loss of generality one can consider only one of these two 
possibilities, say $pppe$).

\begin{table}[h!]
\centering
\caption{Eight different cases of kink tail configurations. Here e denotes an 
exponential tail and p denotes a power law tail (see text for details).}
\vskip 0.3 truecm 
\begin{tabular}{ cccc }
 \hline 
 K$_{1L}$ & K$_{1R}$ & K$_{2L}$ & K$_{2R}$ \\
 \hline
  e & e & e & e \\
  e & p & p & e \\
  p & e & e & p \\
  p & p & p & p \\
  e & e & e & p \\ 
  p & e & e & e \\ 
  p & p & p & e \\ 
  e & p & p & p \\
\hline 
\end{tabular}
\end{table} 

In the five sections III, IV, V, VII and VIII we present one-parameter family 
of potentials 
where the kink tails have the form mentioned in Table 1 with at least one tail 
being a power law tail. Besides, in Sec. VI we present a one-parameter family
of potentials with kink tails of the form $ette$ and $pttp$ where $t, e, p$
correspond to power-tower \cite{KS20}, exponential and power law tail, 
respectively. For simplicity we have omitted all inessential factors appearing 
in these family of potentials. The readers can of course get all these factors 
in the relevant papers cited at appropriate places.

\section{Potentials admitting kink and mirror kink solutions with tails of 
the form $e~p~p~e$}

In this section we present a one-parameter family of potentials of the form
\cite{KS19}
\be\label{3.1}
V(\phi) = \frac{1}{2} \phi^{2n+2}(1-\phi^2)^2  \,, ~~~ n = 1, 2, 3, ... ~.
\ee 
The potentials for $n=1$ and $n=2$ with three degenerate minima are depicted in Fig. 1 
and that for $n=3$ and $n=4$ in Fig. 2. 
This family of potentials has received wide attention in the literature starting with the 
1979 paper of Lohe \cite{Lohe79} where in the context of massless mesons he 
introduced this potential for the case of $n = 1$. The implicit kink solution for
$n = 1, 2, 3$ was first discussed in \cite{KCS}, who pointed out that the kink
tail around $\phi = 0$ has a power law fall off. In 2019 Manton \cite{Manton19} 
attacked the nontrivial problem of the calculation of the KK and the
K-AK force for this model in case $n = 1$ and subsequently this
calculation was extended to the entire family \cite{cddgkks}. We will discuss 
the KK and the K-AK force calculations in detail in Sec. X. 

The potential in Eq. (\ref{3.1}) for any integer $n$ has degenerate minima at 
$\phi = 0, \pm 1$ with $V(\phi = 0, \pm 1) = 0$ and admits a kink solution from 
$0$ to $1$ and a mirror kink solution from $-1$ to $0$ and the corresponding
two antikink solutions with a power law tail around $\phi = 0$ and an exponential 
tail around $\phi = \pm 1$. Unfortunately, in none of these cases, explicit kink 
solutions can be obtained and we can only find implicit kink solutions. From 
the latter we can obtain how a kink profile falls off as $x\rightarrow\pm\infty$.  
It turns out that the nature of the implicit kink solution crucially depends
on whether $n$ is an odd or an even integer. 
We therefore consider the two cases of odd $n=1$ (i.e. potentials of the 
form $\phi^{8}$) and $n=2$ (i.e. potentials of the form $\phi^{10}$) 
separately and then generalize to arbitrary $n$.

\begin{figure}[h] 
\includegraphics[width=6.0 in]{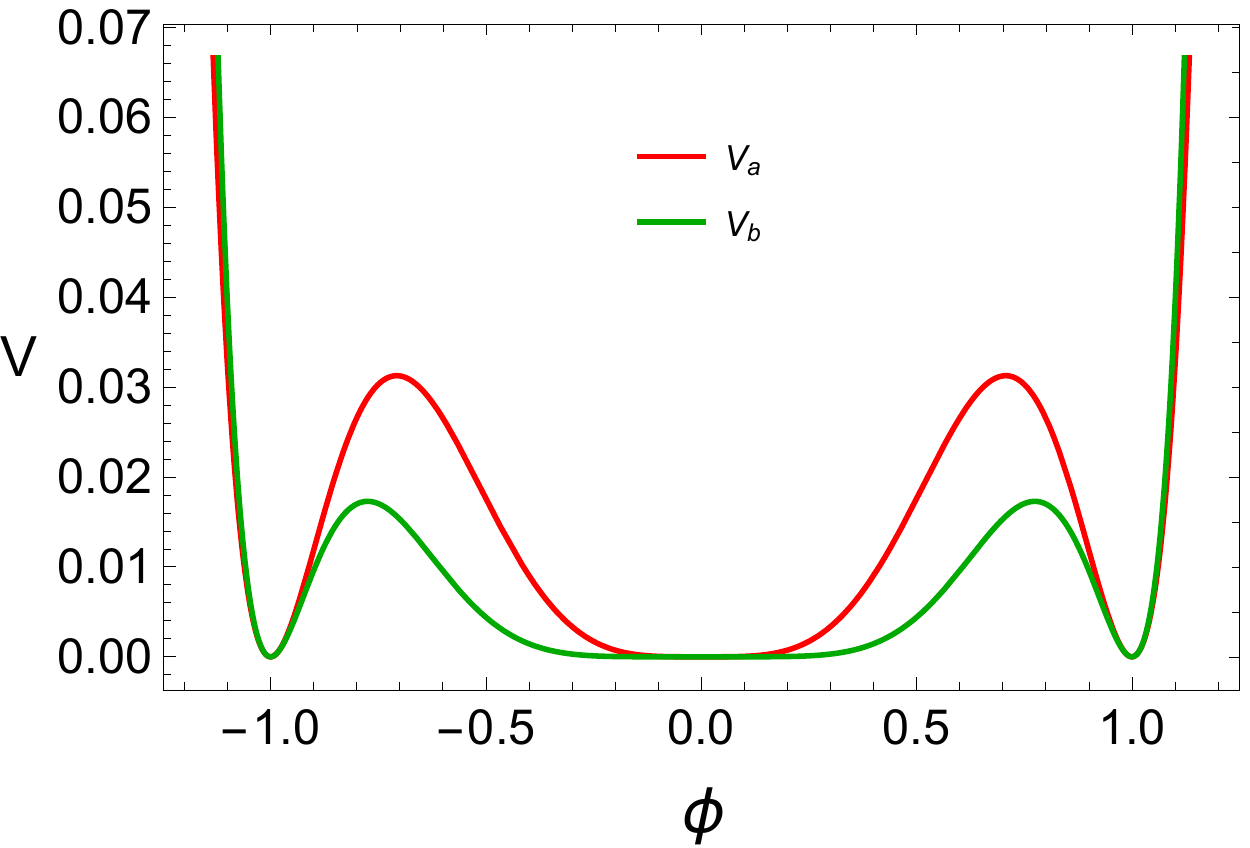}
\caption{Potentials given by Eq. (32) with $n=1$ ($V_a$) and $n=2$ ($V_b$). }
\end{figure} 

\begin{figure}[h] 
\includegraphics[width=6.0 in]{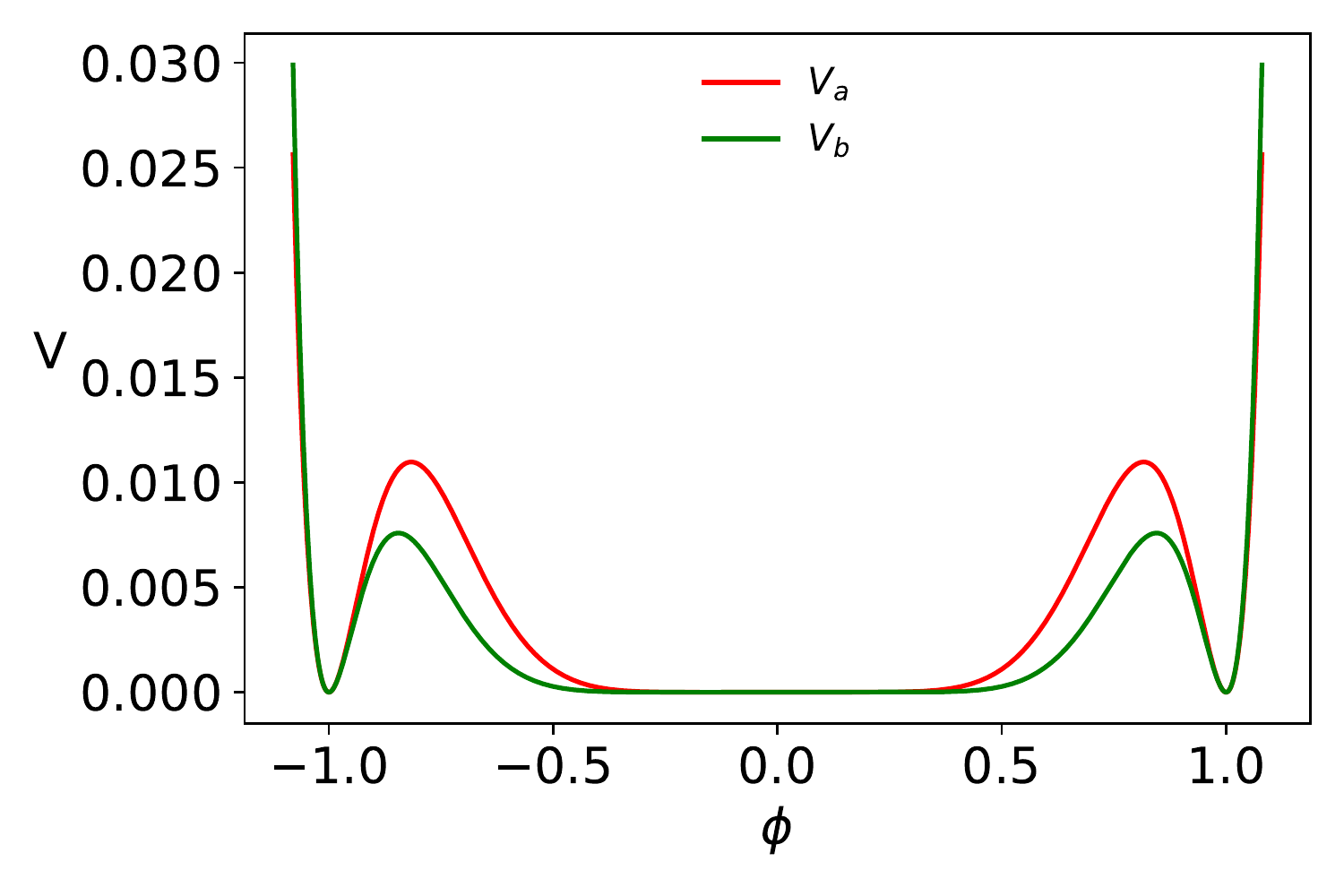}
\caption{Potentials given by Eq. (32) with $n=3$ ($V_a$) and $n=4$ ($V_b$). }
\end{figure} 

\subsection{Case I: $n=1$} 

On using Eq. (\ref{2.3}), the self-dual first order equation for the kink 
solution from $\phi = 0$ to $\phi = 1$ is
\be\label{3.3} 
\frac{d\phi}{dx} =  \phi^2 (1 - \phi^2)\,.  
\ee 
This is easily integrated with the implicit kink solution 
\be\label{3.4} 
 x-A = \frac{1}{\phi} + \frac{1}{2}\ln\frac{1+\phi}{1-\phi} \,,  
\ee 
where $A$ is a constant which without any loss of generality we can put equal
to zero. It is straightforward to show that in case $A = 0$, asymptotically, 
\be\label{3.5} 
\lim_{x \rightarrow -\infty} \phi(x) \simeq \frac{1}{-x} \,, 
~~~ \lim_{x \rightarrow \infty} \phi(x) \simeq 1 - 2 e^{-2 x-2} \,. 
\ee 
Thus the kink tail around $\phi = 0$ is entirely determined by the 
first term on the right hand side of Eq. (\ref{3.4}), i.e. the term 
${1}/{\phi}$.

\subsection{Case II: $n=2$}   

On using Eq. (\ref{2.3}) the self-dual first order equation for the $n = 2$ 
case is
\be\label{3.6} 
\frac{d\phi}{dx} = \phi^3(1-\phi^2) \,. 
\ee
This is easily integrated with the implicit kink solution 
\be\label{3,7} 
2 x = -\frac{1}{\phi^2} + \ln\frac{\phi^2}{1-\phi^2} \,,
\ee 
so that asymptotically, 
\be\label{3.8} 
\lim_{x \rightarrow -\infty} \phi(x) \simeq \frac{1}{\sqrt{-2 x}} \,, 
~~~ \lim_{x \rightarrow +\infty} \phi(x) \simeq 1 - e^{-2 x-1} \,. 
\ee 
Thus the kink tail around $\phi = 0$ is again entirely determined by the 
first term on the right hand side of Eq. (\ref{3,7}), i.e. the term $1/\phi^2$.

Generalization of these results for arbitrary $n$ is straightforward 
\cite{KS19} and one finds that for the 
one-parameter family of potentials as given by Eq. (\ref{3.1}), for arbitrary
integer $n$, while the kink tail falls off like $e^{-2 x}$ around $\phi = 1$, 
it falls off like $x^{-1/n}$ around $\phi = 0$. 

\subsection{Kink Mass}

Using Eq. (\ref{2.5}) one can immediately estimate the kink mass for the
entire family of potentials as given by Eq. (\ref{3.1}). We find that 
\be\label{3.9}
M_K = \frac{2}{(n+2)(n+4)}\,,~~~~n = 1, 2, 3, ...\,,
\ee
so that the kink mass decreases as $n$ increases.
For example, while $M_K(n = 1) = 1/15$, $M_K(n=2)= 1/24$.
Note that the mass of the kink, mirror kink and the two antikinks is the same. 

\section{Potentials admitting kink and mirror kink solutions with tails of 
the form $p~e~e~p$}

In this section we present a one-parameter family of potentials of the form
\cite{KS19}
\be\label{4.1}
V(\phi) = \frac{1}{2} \phi^{2}(1-\phi^2)^{2n+2}  \,, ~~~ n = 1, 2, 3, ... ~.
\ee
The potentials for $n=1$ and $n=2$ with three degenerate minima are depicted 
in Fig. 3 and that for $n=3$ and $n=4$ in Fig. 4. Specifically, these potentials 
have degenerate minima at $\phi = 0\,, \pm 1$ with  $V(\phi = 0, \pm 1) = 0$ and admit 
a kink from $0$ to $1$ and a mirror kink from $-1$ to $0$ and corresponding 
two antikinks. Here while around $\phi = \pm 1$ one has a power law 
tail, around $\phi = 0$ one has an exponential tail.
In these cases too explicit analytic 
kink solutions are not possible and we can only 
find implicit kink solutions. However, from the latter we can determine 
how the kink profile falls off as $x\rightarrow\pm\infty$.  

We will first discuss  the case $n=1$ (i.e. the $\phi^{10}$ field theory) 
\cite{KCS} and $n = 2$ (i.e. $\phi^{14}$ 
field theory) and then mention the behavior of 
the kink tail for arbitrary $n$. 

\begin{figure}[h] 
\includegraphics[width=6.0 in]{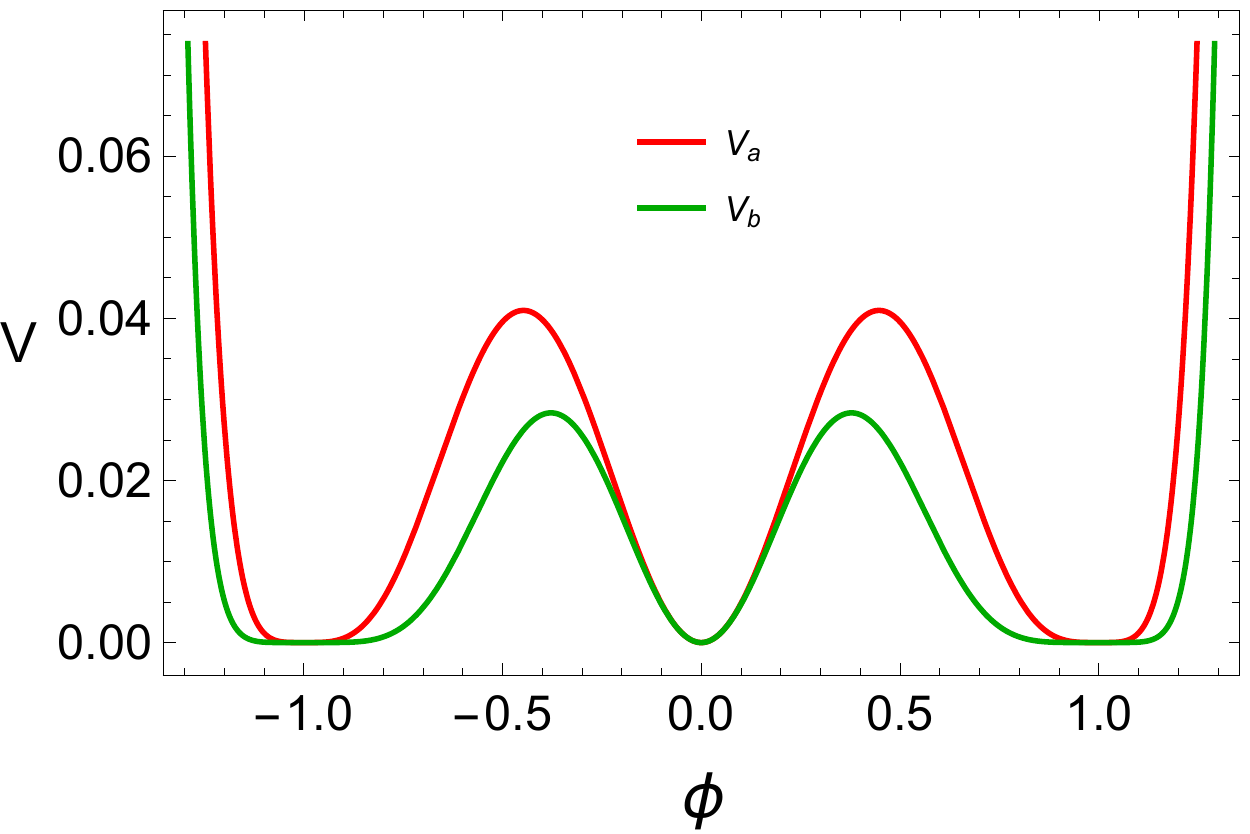}
\caption{Potential given by Eq (40) with $n=1$ ($V_a$) and $n=2$ ($V_b$). }
\end{figure} 

\begin{figure}[h] 
\includegraphics[width=6.0 in]{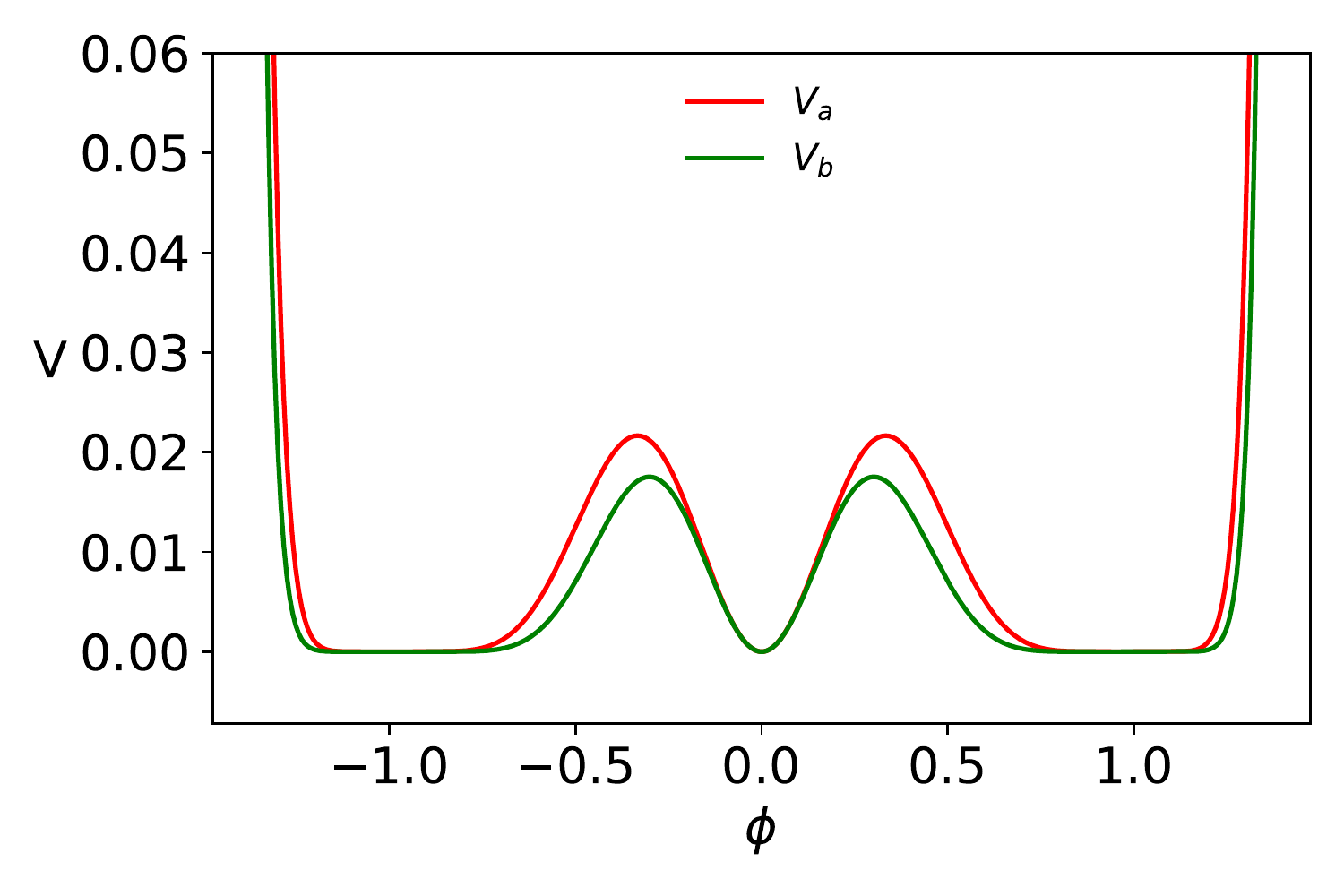}
\caption{Potential given by Eq (40) with $n=3$ ($V_a$) and $n=4$ ($V_b$). }
\end{figure} 

\subsection{Case I: $n=1$}   

On using Eq. (\ref{2.3}), the self-dual first order equation is 
\be\label{4.2} 
\frac{d\phi}{dx} = \phi (1 - \phi^2)^2\,. 
\ee 
This is easily integrated with the implicit kink solution 
\be\label{4.3} 
2 x = \frac{1}{1 - \phi^2} +\ln\frac{\phi^2}{1-\phi^2}\,. 
\ee 
It then follows that asymptotically, 
\be\label{4.4} 
\lim_{x \rightarrow -\infty} \phi(x) \simeq  e^{x-1/2} \,, 
~~~ \lim_{x \rightarrow\infty} \phi(x) \simeq 1 - \frac{1}{4 x} \,. 
\ee 
Thus the kink tail around $\phi = 1$ is entirely determined by the first term on 
the right hand side of the Eq. (\ref{4.3}), i.e. the term $1/(1-\phi^2)$.

\subsection{Case II: $n=2$}   

On using Eq. (\ref{2.3}), the self-dual first order equation for the $n = 2$ 
case is 
\be\label{4.5} 
\frac{d\phi}{dx} = \phi (1 - \phi^2)^3\,. 
\ee 
This is easily integrated with the implicit kink solution 
\be\label{4.6}
2 x = \frac{1}{2(1-\phi^2)^2}+\frac{1}{(1-\phi^2)}
+ \ln\frac{\phi^2}{(1-\phi^2)}\,.  
\ee 
Asymptotically, 
\be\label{4.7} 
\lim_{x \rightarrow -\infty} \phi(x)  \simeq e^{x- 3/4} \,, ~~~
\lim_{x \rightarrow \infty} \phi(x) \simeq 1-\frac{1}{4(x)^{1/2}} \,.  
\ee 
Thus the kink tail around $\phi = 1$ is again entirely determined by the 
first term on the right hand side of Eq. (\ref{4.6}), i.e. the term $1/2(1-\phi^2)^2$.

Generalization of these results for arbitrary $n$ is straightforward 
\cite{KS19} and one finds that for the 
one-parameter family of potentials as given by Eq. (\ref{4.1}), as 
$x \rightarrow -\infty$, the kink tail around $\phi = 0$ falls off like $e^{x}$ 
while as $x \rightarrow +\infty$, the kink tail around $\phi = 1$ falls off 
like $x^{-1/n}$.

\subsection{Kink Mass}

Using  Eq. (\ref{2.5}) one can immediately estimate the kink mass for the
entire family of potentials as given by Eq. (\ref{4.1}). We find that 
\be\label{4.8}
M_K = \frac{1}{2(n+2)}\,,~~~~n = 1, 2, 3, ...\,.
\ee
Thus while $M_K(n = 1) = 1/6, M_K(n=2) = 1/8$, i.e. the kink mass decreases
as $n$ increases.

\section{Potentials admitting kink and mirror kink solutions with tails of 
the form $p~p~p~p$}

In this section we discuss a two-parameter family of potentials  
\be\label{5.1}
V(\phi) = \frac{1}{2} \phi^{2m+2} (1-\phi^2)^{2n+2}  \,, 
~~~ n, m = 1, 2, 3, ... ~\,.
\ee
The potentials for three different cases (i) $m=n=1$, (ii) $m=1, ~n=2$ and (iii) 
$m=2, ~n=1$ each with three degenerate minima are depicted in Fig. 5. 
Similarly, the potentials for three other cases (iv) $m=n=2$, (v) $m=3, ~n=2$ and 
(vi) $m=2, ~n=3$ are depicted in Fig. 6. Specifically, these potentials have 
degenerate minima at $\phi = 0, \pm 1$ and $V(\phi = 0, \pm 1) = 0$ as well as 
admit a kink solution from $0$ to $1$ and a mirror kink solution from $-1$ to 
$0$ and the corresponding antikink solutions, and all of them have 
a power law tail at both the ends. We  look for a kink solution which goes from 
$0$ to +$1$ as $x$ goes from $-\infty$ to +$\infty$, respectively. In these 
cases too the explicit analytic solutions are not possible and we can only 
find implicit kink solutions. From the latter we can obtain 
how a kink profile falls off as $x\rightarrow\pm\infty$.  

In these models depending on if $m < n$ ($m > n$), one can have kink
solutions for which the power law tail around $\phi = 1$ has slower 
(faster) asymptotic fall off compared to the power law tail around $\phi = 0$
while for $m = n$ the power law tails around both $\phi = 0$ and $\phi = 1$ 
have similar fall off. As an illustration we discuss one case each of the three
types.

\begin{figure}[h] 
\includegraphics[width=6.0 in]{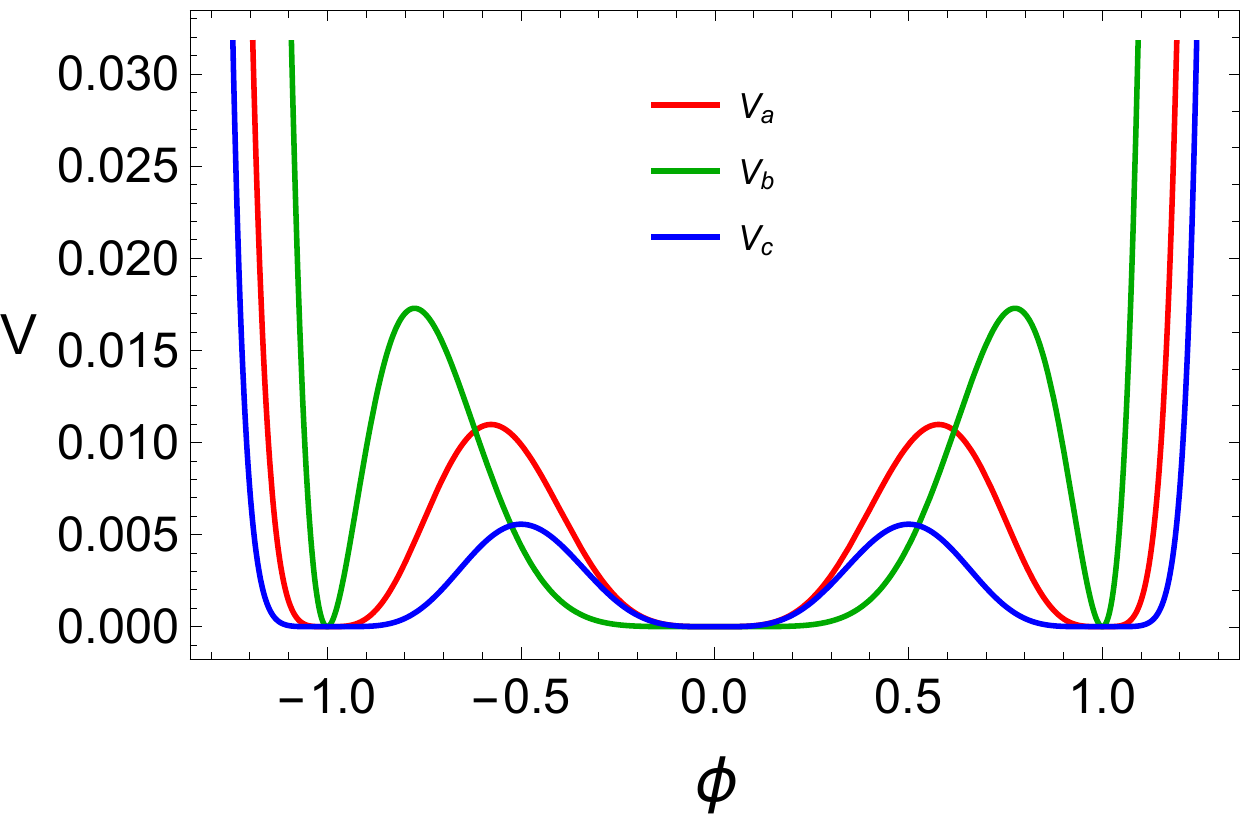}
\caption{Potential given by Eq. (48) with $n=m=1$ ($V_a$), 
$n=1$ and $m=2$ ($V_b$), $n=2$ and $m=1$ ($V_c$). }
\end{figure} 

\begin{figure}[h] 
\includegraphics[width=6.0 in]{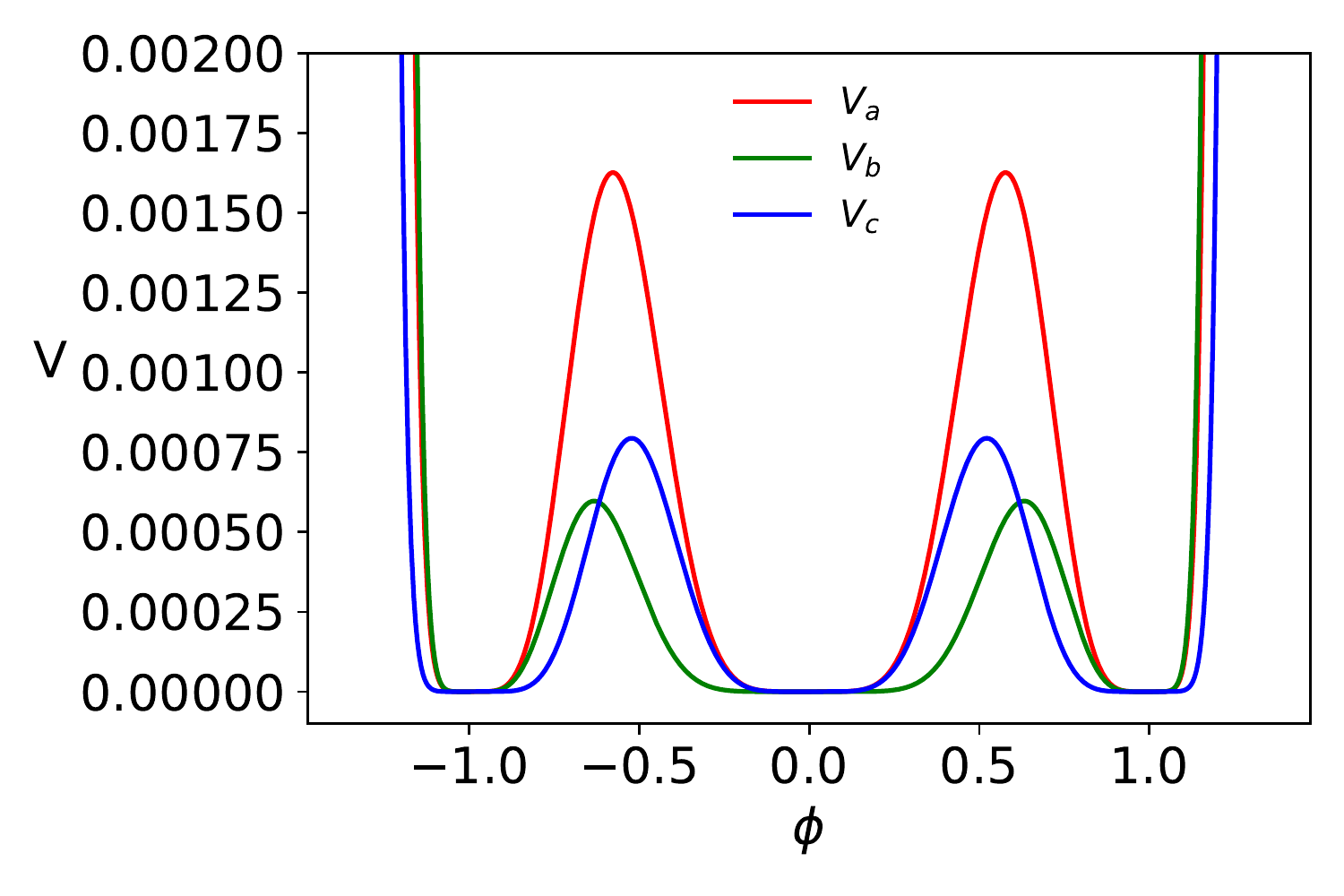}
\caption{Potential given by Eq. (48) with $n=m=2$ ($V_a$), 
$n=2$ and $m=3$ ($V_b$), $n=3$ and $m=2$ ($V_c$). }
\end{figure} 

\subsection{Models where the Kink Tail Around $\phi = \pm 1$ has a Slower 
Asymptotic Fall off Compared to the Tail Around $\phi = 0$, i.e. $m < n$}

For illustration, let us consider the simplest case of $m = 1, n = 2$ in the
potential given by Eq. (\ref{5.1}), i.e.
\be\label{5.2}
V(\phi) = \frac{1}{2} \phi^{4}(1-\phi^2)^{6}\,.
\ee
On using Eq. (\ref{2.3}) the self-dual first order equation is
\be\label{5.3} 
\frac{d\phi}{dx} = \phi^2 (1-\phi^2)^3\,. 
\ee 
This is easily integrated using the identity
\be\label{5.3a}
\int \frac{d\phi}{(1-\phi^2)^{n}} = \frac{(2n-3)}{2(n-1)} 
\int \frac{d\phi}{(1-\phi^2)^{n-1}} + 
\frac{\phi}{2(n-1)(1-\phi^2)^{n-1}}\,,  
\ee 
leading to the implicit kink solution 
\be\label{5.4}
2 x = \frac{\phi}{4(1-\phi^2)^2} -\frac{1}{\phi} +
\frac{7 \phi}{8(1-\phi^2)} +\frac{15}{16} \ln\frac{1+\phi}{1-\phi}\,.
\ee 
Hence asymptotically, 
\be\label{5.5} 
\lim_{x \rightarrow -\infty} \phi(x) \simeq \frac{1}{-2x}\,, 
~~~ \lim_{x \rightarrow\infty} \phi(x) \simeq 1 - \frac{1}{4 \sqrt{2x}}\,. 
\ee 
 Thus the kink tail around $\phi = 1$ is entirely determined by the 
first term on the right hand side of Eq. (\ref{5.4}). 
On the other hand, the kink tail around $\phi = 0$ is entirely determined by 
the second term on the right hand side of Eq. (\ref{5.4}), i.e. the term $1/\phi$.   

\subsection{Models where the Kink Tail Around $\phi = 0$ has a Slower Asymptotic 
Fall off Compared to the Tail Around $\phi = 1$, i.e. $n < m$}

For illustration, let us consider the simplest case of $m = 2, n = 1$ in 
Eq. (\ref{5.1}), i.e. consider the potential
\be\label{5.6} 
V(\phi) = \frac{1}{2} \phi^{6}(1-\phi^2)^4\,.  
\ee
On using Eq. (\ref{2.3}) the self-dual first order equation is
\be\label{5.7} 
\frac{d\phi}{dx} = \phi^3(1-\phi^2)^2 \,.  
\ee
This is easily integrated leading to the implicit kink solution 
\be\label{5.8} 
2x = -\frac{1}{\phi^2} +\frac{1}{1-\phi^2} 
+2 \ln\frac{\phi^2}{1-\phi^2}\,. 
\ee 
Thus asymptotically, 
\be\label{5.9} 
\lim_{x \rightarrow -\infty} \phi(x) \simeq \frac{1}{\sqrt{-2x}} \,, 
~~~ \lim_{x \rightarrow\infty} \phi(x) \simeq 1 - \frac{1}{4 x} \,. 
\ee 
Hence the kink tail around $\phi = 0$ is entirely determined by the 
first term on the right hand side of Eq. (\ref{5.8}). On the other hand, the 
kink tail around $\phi = 1$ is entirely determined by the second term on the 
right hand side of Eq. (\ref{5.8}), i.e. the term $1/(1-\phi^2)$. 

\subsection{Models where the Kink Tails Around $\phi = 0$  and $\phi = 1$ have 
Similar Asymptotic Behavior, i.e. $m = n$}

For illustration, let us consider the simplest case of $m = 1, n = 1$ in 
Eq. (\ref{5.1}), i.e. consider the potential
\be\label{5.10} 
V(\phi) = \frac{1}{2} \phi^{4}(1-\phi^2)^4\,. 
\ee
On using Eq. (\ref{2.3}) the corresponding first order self-dual  
equation is 
\be\label{5.11} 
\frac{d\phi}{dx} =  \phi^2 (1 - \phi^2)^2\,. 
\ee 
This is easily integrated using the identity (\ref{5.3a}) leading to the 
implicit kink solution 
\be\label{5.12} 
x = \frac{\phi}{2(1 - \phi^2)} -\frac{1}{\phi} 
+\frac{3}{4}\ln\frac{1+\phi}{1-\phi}\,. 
\ee 
Thus asymptotically, 
\be\label{5.13} 
\lim_{x \rightarrow -\infty} \phi(x) \simeq \frac{1}{-x}\,, 
~~~ \lim_{x \rightarrow \infty} \phi(x) \simeq 1 - \frac{1}{4x} \,. 
\ee 
Hence, while the kink tail around $\phi = 1$ is entirely determined by the 
first term on the right hand side of Eq. (\ref{5.12}), the kink tail around 
$\phi = 0$ is entirely decided by the second term on the right hand side of 
Eq. (\ref{5.12}), i.e. the term $1/\phi$. As expected, in this case the kink tail 
falls off like $x^{-1}$ around both $\phi = 1$ and $\phi =0$. 

Generalization of these results to the most general potential (\ref{5.1}) with
arbitrary $m$ and $n$ is straightforward and one can show that \cite{KS19} 
for the kink solution between $0$ and $1$ 
the kink tail asymptotically behaves as $x^{-1/m}$ around $\phi = 0$ and as
$x^{-1/n}$ around $\phi = 1$.

\subsection{Kink Mass}

Using Eq. (\ref{2.5}) one can immediately estimate the kink mass for the
entire family of potentials as given by Eq. (\ref{5.1}). We find that 
\be\label{5.14}
M_K = \frac{\Gamma(1+m/2)\Gamma(n+2)}{2\Gamma(n+3+m/2)}\,,
~~~~n, m = 1, 2, 3, ...\,.
\ee
It is easy to check that the kink mass decreases as either $n$ or $m$ increases.  

\subsection{Models Having a Single Kink and an Antikink With Power Law Tails}

Before completing this section it is worth mentioning that the simplest models
admitting kink solution with a power law tail at both the ends are
\be\label{5.15}
V(\phi) = \frac{1}{2} (1-\phi^2)^{2n+2}\,,~~n = 1, 2, 3,... \,.
\ee
The potentials with $n=1$ and $n=2$ are discussed in Section XI (Eq. (146)) and 
shown there  in Fig.~10. These models for arbitrary $n \ge 1$ admit a kink solution 
from $-1$ to $1$ and an antikink solution from $1$ to $-1$ with power law tails 
around both $\phi = \pm 1$. Note that for $n = 0$ we have the celebrated $\phi^4$ 
potential with exponential tails around both $\phi = \pm 1$. Thus these models
for $n \ge 1$ are the simplest generalizations of the $\phi^4$ model but the
tail behavior is entirely different than that for the $\phi^4$ kink. It is also
worth noting that unlike the $\phi^4$ case, for any $n \ge 1$, one can only obtain 
an implicit kink solution from which one can obtain the behavior of the kink
tails around both $\phi = \pm 1$. As an illustration, we first discuss the
simplest case of $n = 1$ and then generalize the results for arbitrary $n$.

\subsubsection{Case I: $n=1$}

On using Eq. (\ref{2.3}), the self-dual first order 
equation is 
\be\label{5.16} 
\frac{d\phi}{dx} = (1 - \phi^2)^2\,. 
\ee 
This is easily integrated using the identity (\ref{5.3a}) and We find
\be\label{5.17} 
4 x = \frac{2\phi}{1-\phi^2} + \ln\left(\frac{1+\phi}{1-\phi}\right) \,. 
\ee 
From here it is straightforward to show that 
\be\label{5.18} 
\lim_{x \rightarrow -\infty} \phi(x) \simeq - 1 + \frac{1}{-4 x} + ... \,.  
~~~ \lim_{x \rightarrow +\infty} \phi(x) \simeq 1 - \frac{1}{4 x} + ... \,. 
\ee 
Note that the leading contribution as $x\rightarrow\pm\infty$ comes from 
the first term on the right hand side of Eq. (\ref{5.17}). 
The generalization to arbitrary $n$ is now straightforward and we find 
that the kink tails around both $\phi = +1$ and $-1$ go like $x^{-1/n}$.

\subsubsection{Kink Mass}

Using Eq. (\ref{2.5}) one can immediately estimate the kink mass for the
entire family of potentials as given by Eq. (\ref{3.1}). We find that 
\be\label{5.19}
M_K = \frac{2^{4n+1} [(2n)!]^2}{(4n+1)!}\,.
\ee
Note again that the kink mass decreases as $n$ increases.

\section{Models with Power-Tower Kink Tails}

In this section we consider two different models giving rise to kink solutions
with kink tails of the form $ette$ and $pttp$, respectively. Here $t$ 
corresponds to the power-tower type of kink tail while $e$ and $p$ as before
correspond to exponential and power law tails, respectively. 

\subsection{Models with Tails of the Form $ette$}

Let us consider a one-parameter family of logarithmic potentials of the form \cite{KS20}
\be\label{6.1}
V(\phi) = (1/2) \phi^{2m+2} [(1/2) \ln(\phi^2)]^2\,,~~m \ge 1\,.
\ee 
The potential for $m=1$ with three degenerate minima is depicted in Fig. 7 and 
for $m=2$ in Fig. 8. These potentials have degenerate minima at $\phi = 0, \pm 1$ 
with $V(\phi = 0, \pm 1) = 0$ while they have degenerate maxima at
\be\label{6.2}
\phi_{max} = \pm \, e^{-1/(m+1)}\,,~~~V_{max} = \frac{1}{2e^2 (m+1)^2}\,.
\ee
Thus notice that while $\phi_{max}(m = 1) = \pm e^{-1/2}$, as $m$ becomes
larger, $\phi_{max}$ moves towards $\pm 1$. On the other hand 
while, $V_{max}(m =1) = {1}/{8e^2}$, as $m$ becomes larger, $V_{max}$ 
decreases progressively towards zero. All these models for any integer $m$  
admit a kink solution from $0$ to $1$ and a mirror kink solution from $-1$ to 
$0$ (and the corresponding antikink solutions) with exponential tails around 
$\phi = \pm 1$ and power tower tails around $\phi = 0$. 

For the potential (\ref{6.1}) we need to solve the self-dual first order 
equation  
\be\label{6.3}
\frac{d\phi}{dx} = \pm \phi^{m+1} [(1/2) \ln(\phi^2)]\,.
\ee
For the kink solution between $0$ and $1$ we need to solve the self-dual 
Eq. (\ref{2.5}) with negative sign.  
This is easily integrated by making the substitution $t = (1/2)\ln(\phi^2)$
and we obtain the implicit kink solution
\be\label{6.4}
-x = \int \frac{e^{-mt}}{t}\, dt = Ei(-mt)\,,
\ee
where $Ei(x)$ denotes the exponential integral function \cite{erdelyi, grad}. 
Unfortunately, we do not know how to invert this function analytically
\cite{inverse} and obtain $t$ and hence $\phi$ as a function of $x$. However, 
using the Taylor series expansion of $Ei(x)$ \cite{erdelyi}
\be\label{6.5}
Ei(x) = \gamma + \ln |x| + x + \frac{x^2}{2\,2!}+... \,,
\ee
as well as the asymptotic formula \cite{erdelyi}
\be\label{6.6}
Ei(x) = e^{x} \left [\frac{1}{x} + \frac{1}{x^2} + \frac{2!}{x^3}
+\frac{3!}{x^4}+...\right]\,,
\ee
one can estimate the tail behavior around $\phi = 0$ as $x \rightarrow 
-\infty$ and around $\phi = 1$ as $x \rightarrow +\infty$. Here $\gamma
= 0.577$ is Euler's constant. One finds that  
\be\label{6.7}
\lim_{x \rightarrow -\infty} \phi^{m}(x) \ln[\phi(x)] 
\simeq \frac{1}{mx}\,,~~
\lim_{x \rightarrow +\infty} \phi(x) \simeq 1- \frac{e^{-(x+\gamma)}}{m}\,.
\ee 
It is worth pointing out that the asymptotic behavior around $\phi = 0$
(as $x \rightarrow -\infty$) in Eq. (\ref{6.7}) can also be written as 
\be\label{6.8}
\lim_{x \rightarrow -\infty} \phi(x)^{{\phi(x)}^{m}} \simeq e^{1/mx}\,,
\ee
which is known in the literature as the power-tower function of order two 
\cite{powert} or tetration \cite{tetration}.  

If one inverts Eq. (\ref{6.7}) numerically one finds that asymptotically
as $x \rightarrow -\infty$, around $\phi = 0$ the power tower tail
essentially behaves as a power law tail where the exponent is not known
precisely.

\begin{figure}[h] 
\includegraphics[width=6.0 in]{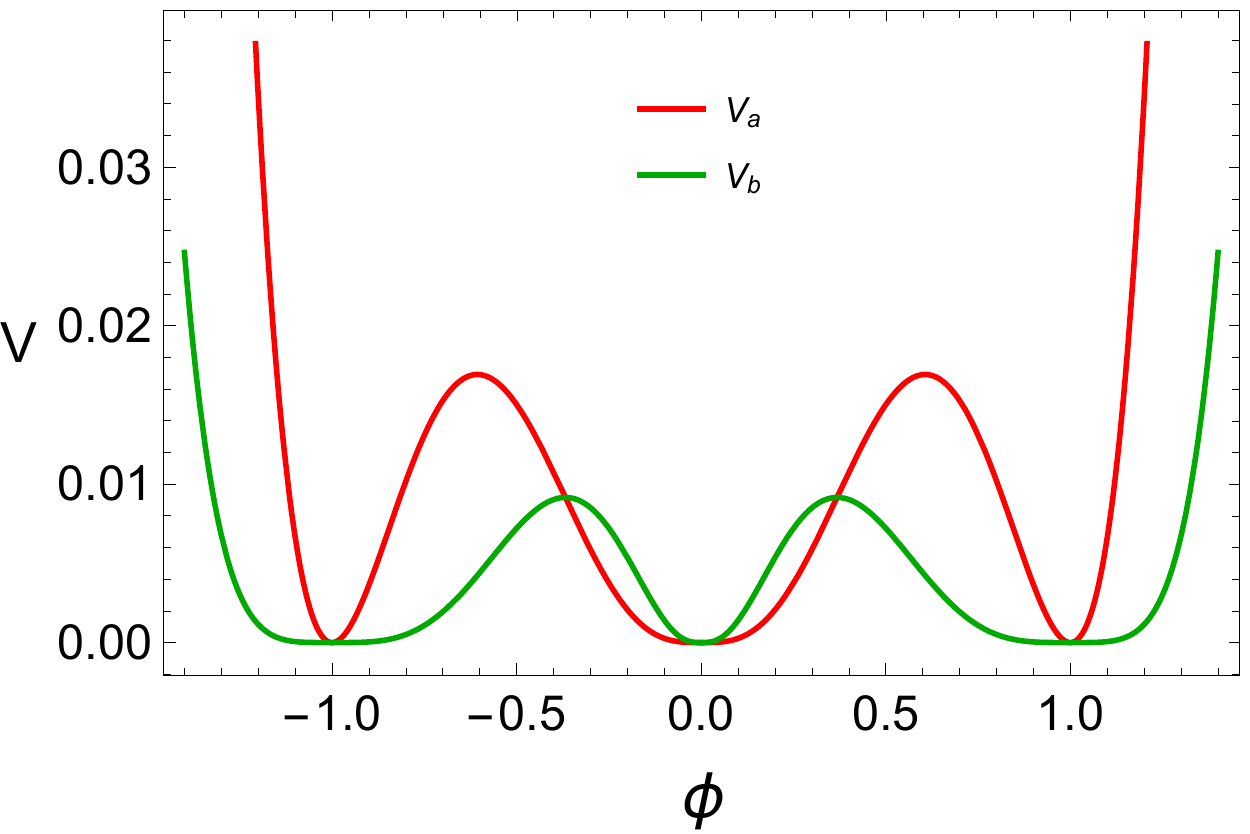}
\caption{Potential given by Eq. (68) with $m=1$ ($V_a$) and Eq. (77) with 
$n=1$ and $m=1$ ($V_b$). }
\end{figure} 

\begin{figure}[h] 
\includegraphics[width=6.0 in]{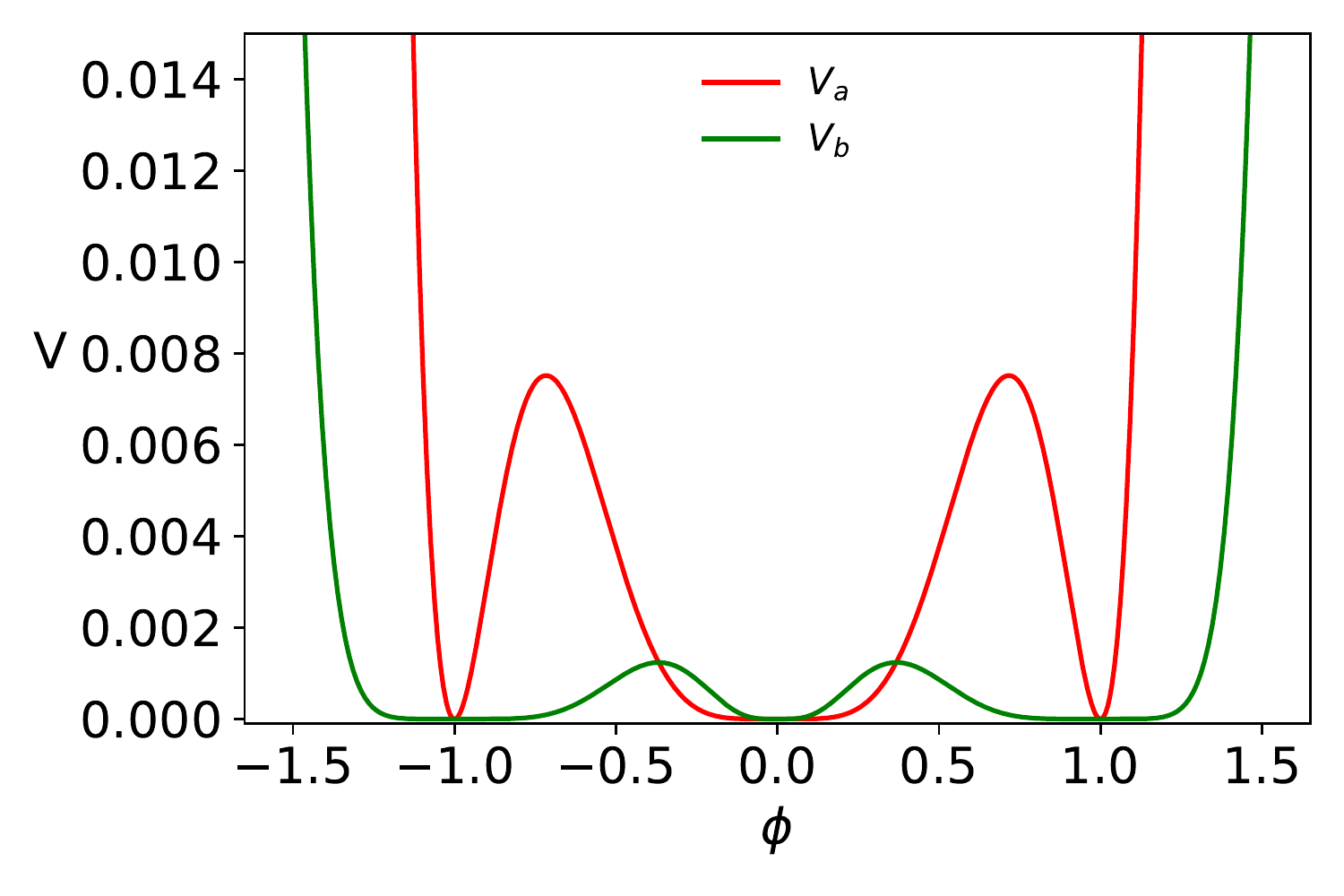}
\caption{Potential given by Eq. (68) with $m=2$ ($V_a$) and Eq. (77) with 
$n=2$ and $m=2$ ($V_b$). }
\end{figure} 

\subsubsection{Kink Mass}

One can easily calculate the kink mass for the entire family of potentials. 
In particular, for the kink potential as given by Eq. (\ref{6.1}), the kink 
mass turns out to be 
\be\label{6.9}
M_K = \frac{1}{(m+2)^{2}}\,.  
\ee
Observe that even in this case the kink mass decreases as $m$ increases. 
 

\subsection{Models with Tails of the Form $pttp$}

Let us consider a two-parameter family of logarithmic potentials  
\be\label{6.10}
V(\phi) = (1/2) \phi^{2m+2} [(1/2) \ln(\phi^2)]^{2n+2}\,,~~m, n \ge 1\,.
\ee 
The potential for $m=1$ and $n=1$ with three degenerate minima is depicted 
in Fig. 7. Similarly, the potential with $m=2$ and $n=2$ is depicted in Fig. 8. 
These potentials have degenerate minima at $\phi = 0, \pm 1$ with 
$V(\phi = 0, \pm 1) = 0$ while they have degenerate maxima at
\be\label{6.11}
\phi_{max} = \pm \, e^{-(n+1)/(m+1)}\,,~~~
V_{max} = \frac{1}{2e^{2(n+1)}} \left[\frac{(n+1)}{(m+1)}\right]^{2(n+1)}\,.
\ee
Notice that both $\phi_{max}$ and $V_{max}$ depend on two parameters $m$ and
$n$. Further, for a fixed $m$, as $n \rightarrow \infty$, $\phi_{max}
\rightarrow 0$ and $V_{max} \rightarrow \infty$. On the other hand, for a fixed
$n$, as $m \rightarrow \infty$, $\phi_{max} \rightarrow 1$ and 
$V_{max} \rightarrow 0$. Finally, for $m = n$, $\phi_{max} = \pm e^{-1}$
and the corresponding $V_{max} = {1}/{2e^{2(n+1)}}$.  
It is interesting to note that for a given $m$, all the potentials as 
given by Eq. (\ref{6.10}) with arbitrary integer $n$ have the same value
$V(\phi) = \frac{1}{2e^{2(m+1)}}$ in case $\phi = \pm 1/e$  or 
$V(\phi) = \frac{e^{2(m+1)}}{2}$ in case $\phi = \pm e$. 

All these models, for any integers $m$  and $n$ admit a kink solution from $0$ 
to $1$ and a mirror kink solution from $-1$ to $0$ (and corresponding antikink 
solutions) with a power law tail around $\phi = \pm 1$ and a power-tower tail 
around $\phi = 0$. 

In order to obtain the kink solution from $0$ to $1$, we need to solve the 
self-dual equation  
\be\label{6.12}
\frac{d\phi}{dx} = \pm \phi^{m+1} \big([(1/2) \ln(\phi^2)]\big)^{n+1}\,.
\ee
This is easily done by making the substitution $t = (1/2)\ln(\phi^2)$
and we obtain \cite{KS20} the implicit kink solution 
\be\label{6.13}
\pm x = -e^{-mt}  \sum_{k=1}^{n} \frac{(-m)^{k-1}}{n(n-1)...(n+1-k) 
t^{n+1-k}} +\frac{(-m)^{n}}{n!} Ei(-mt)\,, 
\ee
where we need to take $+x$ $(-x)$ in Eq. (\ref{6.12}) depending on whether
$n$ is an odd (or even) integer. We then find that   
\bea\label{6.14}
&&\lim_{x \rightarrow -\infty} \phi^{m}(x) \big(\ln[\phi(x)]\big)^{n+1} 
 \simeq \frac{(-1)^{n}}{mx}\,, ~~~
\lim_{x \rightarrow +\infty} \phi(x) \simeq 1
- \frac{1}{\left[nx+\frac{m^{n}\gamma}{(n-1)!}\right]^{1/n}}\,.
\eea

\subsubsection{Kink Mass}

One can easily calculate the kink mass for the entire family of potentials 
given by Eq. (\ref{6.10}) and we find 
\be\label{6.15}
 M_K =  \frac{(n+1)!}{(m+2)^{n+2}}\,.   
\ee
The kink mass decreases as $m$ increases keeping $n$ fixed. On the other hand, 
the kink mass increases (decreases) as $n$ increases keeping $m$ fixed 
depending on the values of $m$ and $n$. For example $M_{K}(n,m) > (<)$  
$M_{K}(n+1,m)$ depending on if $m > (<)$ $n+1$. 

\section{Kink solutions with tails of the form $p~p~p~e$}

We now briefly discuss a one-parameter family of potentials of the form 
\cite{KS19}
\be\label{7.1}
V(\phi) = \frac{1}{2} (1 - \phi^2)^{2n+2} (2-\phi^2)^{2}  \,,
~~~ n = 1, 2, 3, ... ~\,.
\ee
These potentials have degenerate minima at $\phi = \pm 1, \pm \sqrt{2}$ with 
$V(\phi = \pm 1, \pm \sqrt{2}) = 0$ and admit a kink solution from $-1$ to $1$ 
and a kink solution from $1$ to $\sqrt{2}$ as well as a mirror kink solution from 
$-\sqrt{2}$ to $-1$ and the three corresponding antikink solutions. 
While the kink from $-1$ to $1$ has a power law 
tail around both $\phi = -1$ as well as $\phi = 1$, the kink from $1$ to 
$\sqrt{2}$ has a power law tail around $\phi = 1$ and an exponential tail 
around  $\phi = \sqrt{2}$. In these cases too we can only obtain implicit kink 
solutions from which we can obtain the behavior of kink tails. 

\subsection{ $-1$ to $1$ Kink solution}

On using Eq. (\ref{2.3}), the self-dual first order equation for the potential
(\ref{7.1}) is 
\be\label{7.2}
\frac{d\phi}{dx} = (2 -\phi^2) (1-\phi^2)^{n+1}\,. 
\ee 
This is easily integrated with the solution \cite{KS19} 
\be\label{7.3} 
2 x = \frac{\phi}{n (1 -\phi^2)^{n}} 
+ \frac{(-1)^{n+1}}{\sqrt{2}} \ln \left(\frac{\sqrt{2}+\phi}
{\sqrt{2}-\phi}\right) + lower~order~terms\,.
\ee
Note that in Eq. (\ref{7.3}) we have only specified those terms which 
contribute to the dominant asymptotic behavior as $x \rightarrow \pm \infty$.  
Asymptotically, 
\bea\label{7.4} 
&&\lim_{x \rightarrow -\infty} \phi(x) \simeq -1 
+ \bigg [\frac{-1}{n 2^{n+1} x} \bigg ]^{1/n}\,, ~~~
\lim_{x \rightarrow +\infty} \phi(x) \simeq 1
- \bigg [\frac{1}{n 2^{n+1} x} \bigg ]^{1/n}\,.
\eea
Thus for the kink solution from $-1$ to $1$, the kink tail around both 
$\phi = -1$ and $\phi = 1$ has a power law tail going like $x^{-1/n}$.

\subsubsection{Mass of $-1$ to $1$ Kink}

It is straightforward to calculate the kink mass for the entire family
as given by Eq. (\ref{7.1}) and we find that 
\be\label{7.4a}
M_K(-1,1) = 2(2n+9/2) B[3/2,n+2]\,,
\ee
where $B[a,b]$ is Euler's beta function \cite{abs}.

\subsection{$1$ to $\sqrt{2}$ Kink solution}

On using Eq. (\ref{2.3}), the self-dual first order equation is given by 
\be\label{7,5}
\frac{d\phi}{dx} = (2 -\phi^2) (\phi^2 -1)^{n+1} \,. 
\ee 
This is easily integrated with the solution 
\be\label{7.6} 
2 x = \ln \left(\frac{\sqrt{2}+\phi}{\sqrt{2}-\phi}\right) 
- \frac{\sqrt{2} \phi} {n (\phi^2 -1)^{n}} + lower~order~terms\,.
\ee
Note that in Eq. (\ref{7.6}) we have only specified those terms which 
contribute to the dominant asymptotic behavior as $x \rightarrow \pm \infty$.  
Asymptotically, 
\bea\label{7.7} 
&&\lim_{x \rightarrow -\infty} \phi(x) \simeq 1 
+ \bigg [\frac{-\sqrt{2}}{n 2^{n+1}x} \bigg ]^{1/n}\,,  ~~~
\lim_{x \rightarrow +\infty} \phi(x) \simeq \sqrt{2} - h(n) e^{-2 x}\,, 
\eea 
where $h(n)$ is a known function of $n$. Thus for the kink solution from $1$ 
to $\sqrt{2}$, while one has an exponential tail around $\phi = \sqrt{2}$, the 
kink tail around $\phi = 1$ goes like $(-x)^{-1/n}$ as $x \rightarrow  -\infty$. 

\subsubsection{Mass of $1$ to $\sqrt{2}$ Kink}

It is straightforward to calculate the kink mass for the entire family
as given by Eq. (\ref{7.1}) and we find that 
\be\label{7.8}
M_K(1,\sqrt{2}) = \frac{1}{2\sqrt{2}(n+2)(n+4)}\, _{2}F_{1}(1/2, 2, n+4, 1/2)\,,
\ee
where $_2F_1$ denotes a hypergeometric function. 

\section{Kink Solutions with Tails of the Form $e~e~e~p$}

We now briefly discuss a one-parameter family of potentials of the form 
\cite{KS19}
\be\label{8.1}
V(\phi) = \frac{1}{2} (1 - \phi^2)^2 (2-\phi^2)^{2n+2} \,, 
~~~ n = 1, 2, 3, ... ~\,.
\ee
These potentials have degenerate minima at $\phi = \pm 1, \pm \sqrt{2}$ with
$V(\phi = \pm 1, \pm \sqrt{2}) = 0$ and admit a kink solution from $-1$ to $1$ 
and a kink solution from $1$ to $\sqrt{2}$ as well as a mirror kink solution from 
$-\sqrt{2}$ to $-1$ and the corresponding three antikink solutions. While the 
kink from $-1$ to $1$ has an exponential tail 
around both $\phi = \pm 1$, the kink from $1$ to $\sqrt{2}$
has an exponential tail around $\phi = 1$ and a power law tail around 
$\phi = \sqrt{2}$. In these cases too one can only find implicit kink solutions 
from which one can obtain the various kink tails. 

\subsection{ $-1$ to $1$ kink solution}

On using Eq. (\ref{2.3}), the self-dual first order equation for the potential
(\ref{8.1}) is 
\be\label{8.2}
\frac{d\phi}{dx} = (1 -\phi^2) (2-\phi^2)^{n+1}\,. 
\ee 
This is easily integrated with the solution \cite{KS19} 
\be\label{8.3} 
2 x = \ln \left(\frac{1+\phi}{1-\phi}\right) - \frac{\phi}
{2n (2 -\phi^2)^{n}} + lower~order~terms\,.
\ee
Note that in Eq. (\ref{8.3}) we have only specified those terms which 
contribute to the dominant asymptotic behavior as $x \rightarrow \pm \infty$.  
Asymptotically we find that, 
\be\label{8.4} 
\lim_{x \rightarrow -\infty} \phi(x) \simeq -1 + f(n) e^{2 x} \,, 
~~~ \lim_{x \rightarrow +\infty} \phi(x) \simeq 1 
- f(n) e^{-2 x} \,,   
\ee 
where $f(n)$ is a known function of $n$. Thus for the kink solution from $-1$ to 
$1$, the kink tail around both $\phi = -1$ and $\phi = 1$ has an exponential tail. 

\subsubsection{Mass of $-1$ to $1$ Kink}

It is straightforward to calculate the kink mass for the entire family
as given by Eq. (\ref{8.1}) and we find that 
\be\label{8.4a}
M_K(-1,1) = \frac{2^{n+3}}{3}\, _{2}F_{1}(-n-1, 1/2, 5/2, 1/2)\,.
\ee

\subsection{ $1$ to $\sqrt{2}$ kink solution}

On using Eq. (\ref{2.3}), the self-dual first order equation is now given by 
\be\label{8.5}
\frac{d\phi}{dx} = (\phi^2 -1) (2-\phi^2)^{n+1}\,.
\ee 
This is easily integrated with the solution \cite{KS19} 
\be\label{8.6} 
2 x = \ln \left(\frac{\phi-1}{\phi+1}\right) + \frac{\phi}
{2 n (2 -\phi^2)^{n}} + lower~order~terms\,.
\ee
Note that in Eq. (\ref{8.6}) we have only specified those terms which 
contribute to the dominant asymptotic behavior as $x \rightarrow \pm \infty$.  
Asymptotically, 
\bea\label{8.7} 
&&\lim_{x \rightarrow -\infty} \phi(x) \simeq 1 - f(n) e^{2 x}\,,  ~~~
\lim_{x \rightarrow +\infty} \phi(x) \simeq \sqrt{2}
- \bigg [\frac{1}{n 2^{2n+1} x} \bigg ]^{1/n} \,. 
\eea 
Thus for the kink solution from $1$ to $\sqrt{2}$, while around 
$\phi = 1$ one has an exponential tail, the kink tail around $\phi = \sqrt{2}$ 
goes like $x^{-1/n}$ as $x \rightarrow  \infty$.

\subsubsection{Mass of $1$ to $\sqrt{2}$ Kink}

It is straightforward to calculate the kink mass for the entire family
as given by Eq. (\ref{8.1}) and we find that 
\be\label{8.8}
M_K(1,\sqrt{2}) = \frac{1}{2\sqrt{2}(n+2)(n+4)}\, 
_{2}F_{1} (1/2, 2, n+4, 1/2)\,.
\ee

\section{Explicit Kink Solutions with Power Law Tails}

In the last six sections we have presented a number of one-parameter family
of potentials wherein one could obtain kink solutions such that at least one of the
kink tails has a power law fall off. Unfortunately, in all these cases one could 
only obtain implicit kink solutions. It is clearly desirable and of interest 
to look for models where one could obtain kink solutions in an explicit form. 
That would offer a deeper insight into the various aspects of kink 
solutions with power law tails. For example, one could then explicitly 
calculate the kink stability potential
as defined in Sec. II (see Eqs. (\ref{2.7}) and (\ref{2.7a})) and
verify that for kink solutions with power law tails, indeed there is no gap 
between the zero mode and the beginning of
the continuum thereby providing a concrete example to the proof given in Sec. 
II. We now discuss three models, two polynomial and one nonpolynomial types 
where explicit kink solutions with power law tails can be obtained. 

\subsection{Model I}

We now obtain explicit kink solutions with power law tails 
in a one-parameter family of potentials characterized by the potential
\cite{KS21}
\be\label{9.1}
V(\phi) = \frac{1}{2} \phi^{2(n+1)} |(1 - \phi^{2n})|^{3}\,.
\ee
Note that this potential has degenerate minima at $ \phi = 0, \pm 1$ with
$V(\phi = 0, \pm 1) = 0$ and admits a 
kink solution from $0$ to $1$ and a mirror kink solution
from $-1$ to $0$ and the corresponding two antikink solutions. 
It is worth pointing out 
that whereas the potential (\ref{9.1}) is continuous, its derivative is 
discontinuous at $\phi = \pm 1$.  However, since for the kink as well for the 
antikink solutions $-1 \le \phi \le 1$, this discontinuity would not matter as 
far as the kink and the antikink solutions are concerned.

In order to obtain the kink solution from 0 to $1$, we need to solve the 
self-dual equation
\be\label{9.2}
\frac{d\phi}{dx} =  \phi^{n+1} |(1-\phi^{2n})|^{3/2}\,.
\ee
This is easily integrated yielding
\be\label{9.3}
\frac{(2\phi^{2n} -1)}{\phi^{n} (1-\phi^{2n})^{1/2}} = n x\,.
\ee
Eq. (\ref{9.3}) is inverted with ease yielding an explicit kink solution
\be\label{9.4}
\phi(x) = \frac{1}{2^{1/2n}}\left[1+\frac{n x}
{\sqrt{n^2 x^2+4}}\right]^{1/2n}\,.
\ee
It is straightforward to see that
\be\label{9.5}
\lim_{x \rightarrow -\infty} \phi(x) \simeq \frac{1}{(-n x)^{1/n}}\,,
~~~\lim_{x \rightarrow \infty} \phi(x) \simeq 1 - \frac{1}{2n(n x)^{2}}\,.
\ee

Since the kink solution is explicitly known, the kink stability potential 
$V(x)$ which appears in the Schr\"odinger-like equation (\ref{2.7})  
is easily calculated using Eqs. (\ref{2.7a}) and (\ref{9.4})
\bea\label{9.6}
&&V(x) = \frac{1}{4(4+n^2 x^2)} \bigg [-(58 n^2+33n-1) 
+\frac{(14n^2+3n+1)n^2 x^2}{(4+n^2 x^2)} 
+\frac{2(10n^2+3n-1)n x}{\sqrt{(4+n^2 x^2)}} \bigg ]\,.
\eea
As expected, this kink potential $V(x)$ vanishes as 
$x \rightarrow \pm \infty$
thereby confirming that indeed in this case there is no gap between the
zero mode and the beginning of the continuum. Further, one finds that
$V(x = 0) = -\frac{(58 n^2+33 n-1)}{16}$.   

Using the explicit kink solution (\ref{9.4}), it is straightforward to 
calculate the translational zero mode in the kink stability Schr\"odinger-like 
Eq. (15)
\be\label{9.7}
\psi_{0} \propto \frac{d\phi(x)}{dx} \propto 
 \frac{1}{[1+\frac{n x}{\sqrt{n^2 x^2+4}}]^{1-1/2n} (1+n^2 x^2)^{3/2}}\,.
\ee
As expected this zero mode vanishes as $x\rightarrow\pm\infty$, i.e. as 
$\phi \rightarrow 0, a$.

\subsubsection{Kink Mass}

It is easy to calculate the kink mass in this case. We find
\be\label{9.9}
M_K = \frac{3 \sqrt{\pi}}{8 n} 
\frac{\Gamma[(n+2)/2n]}{\Gamma[(3n+1)/n]}\,.
\ee

\subsection{Model II}

Let us consider a one-parameter family of potentials \cite{KS21}
\be\label{9.10}
V(\phi) = \frac{1}{2} |(1 - \phi^{2n})|^{(2n+1)/n}\,.
\ee
Note that this potential has degenerate minima at $\pm 1$ with $V(\phi = \pm 1)
= 0$ and admits a kink
solution from $-1$ to $1$ and the corresponding antikink solution from $1$ to 
$-1$. Note that as in the previous example, whereas the 
potential (\ref{9.10}) is continuous, its derivative is discontinuous at 
$x=\pm 1$. However, since for the kink as well as
for the antikink solutions $-1\le \phi \le 1$, this discontinuity would not 
matter as far as the kink and the antikink solutions are concerned.
 
In order to obtain the kink solution from $-1$ to $1$, we need to solve the
self-dual equation 
\be\label{9.11}
\frac{d\phi}{dx} = |1-\phi^{2n}|^{(2n+1)/2n}\,.
\ee
This is easily integrated yielding 
\be\label{9.12}
\frac{\phi}{(1-\phi^{2n})^{1/2n}} = x\,.
\ee
Eq. (\ref{9.12}) is inverted with ease yielding an explicit kink solution
\be\label{9.13}
\phi(x) = \frac{x}{[1+(x)^{2n}]^{1/2n}}\,.
\ee
It is straightforward to see that
\be\label{9.14}
\lim_{x \rightarrow -\infty} \phi(x) \simeq -1 + \frac{1}{2n(x)^{2n}}\,,
~~~\lim_{x \rightarrow \infty} \phi(x) \simeq 1 - \frac{1}{2n(x)^{2n}}\,.
\ee
It is worth pointing out that for the special case of n = 1, the kink solution
(\ref{9.13}) has been obtained previously \cite{Gomes}. 

It is easily checked that, as expected the zero mode eigenfunction vanishes
as $x\rightarrow\pm\infty$, i.e. as $\phi \rightarrow\pm 1$. 
Using the explicit kink solution (\ref{9.13}), the kink stability potential 
$V(x)$ which appears in the 
Schr\"odinger like Eq. (\ref{2.7}) is easily calculated 
\be\label{9.15}
V(x) = (2n+1) \frac{(x)^{2n-2}[(2n+2) x^{2n}-{2n-1}]}{[1+(x)^{2n}]^2}\,.
\ee
As expected, this kink stability potential vanishes as $x\rightarrow\pm\infty$ 
thereby confirming that indeed in this case there is no gap between the zero 
mode and the beginning of the continuum. 

Using the explicit kink solution (\ref{9.13}) it is straightforward to calculate
the zero mode and we find 
\be\label{9.16}
\psi_{0} \propto \frac{d\phi(x)}{dx} \propto [1+(x)^{2n}]^{-(2n+1)/2n}\,.
\ee

\noindent{\bf Kink Mass}

Finally, it is easy to calculate the kink mass in this case. We find
\be\label{9.17}
M_K = \frac{\Gamma[(4n+1)/2n] \Gamma[1/2n]}{n \Gamma[(2n+1)/n]}\,.
\ee

\subsection{Model III}

Let us consider the periodic potential \cite{mra,ko}
\be\label{9.18}
V(\phi) = \frac{1}{2} \cos^4(\phi)\,.
\ee
Note that this periodic potential has degenerate minima at 
$\phi = \pm \pi/2$ with $V(\phi = \pm \pi/2) = 0$ and admits a kink
solution from $-\pi/2$ to $\pi/2$ and the corresponding antikink solution 
from $\pi/2$ to $-\pi/2$. Note that unlike the two previous examples, 
not only the potential (\ref{9.18}) but its derivative is also continuous.
 
In order to obtain the kink solution from $-\pi/2$ to $\pi/2$, we need 
to solve the self-dual equation 
\be\label{9.19}
\frac{d\phi}{dx} = \cos^2(\phi)\,.
\ee
This is easily integrated yielding the kink solution
\be\label{9.20}
\phi(x) = \tan^{-1}(x)\,.
\ee
It is straightforward to see that
\be\label{9.21}
\lim_{x \rightarrow -\infty} \phi(x) \simeq -\pi/2 + \frac{1}{x}\,,
~~~\lim_{x \rightarrow \infty} \phi(x) \simeq \pi/2 - \frac{1}{x}\,.
\ee

Using the explicit kink solution (\ref{9.20}), the kink stability potential 
$V(x)$ which appears in the 
Schr\"odinger like Eq. (\ref{2.7}) is easily calculated 
\be\label{9.22}
V(x) = \frac{6 x^2 -8}{(1+x^2)^2}\,.
\ee
As expected, this kink stability potential vanishes as $x\rightarrow\pm\infty$ 
thereby confirming that indeed in this case there is no gap between the zero 
mode and the beginning of the continuum. 
Using the explicit kink solution (\ref{9.20}) it is straightforward to 
calculate the zero mode and we find 
\be\label{9.23}
\psi_{0} \propto \frac{d\phi(x)}{dx} \propto \frac{1}{(1+x^2)}\,.
\ee

\noindent{\bf Kink Mass}

Finally, it is easy to calculate the kink mass in this case. We find
\be\label{9.24}
M_K = \frac{3\pi}{8}\,. 
\ee

We might add here that apart from the nonpolynomial potential (\ref{9.18})
discussed above, a couple of one-parameter family of nonpolynomial models have
also been introduced \cite{bmm} for which explicit kink solutions have
been obtained.

\section{Kink-Kink and Kink-Antikink Forces}

Now we turn to perhaps the most important but not so well understood topic
of the calculation of the kink-kink (KK) and the kink-antikink (K-AK) forces 
amongst two widely separated kinks with power law tails which are capable of 
interacting over very large distances. This is in contrast to the well known $\phi^4$ 
and many other kinks with exponential tails for which the calculation of the 
KK force between two well separated kinks relies on a linear superposition 
of the exponentially small tails in the region between the kinks. 
One then finds that for such kinks the KK as well as the K-AK forces decay 
exponentially fast with the kink separation. Further, while the KK force is
repulsive and the K-AK force is attractive, the magnitude of the KK and the
K-AK force is equal. 

On the other hand, for the extended kinks (i.e. kinks with a power law tail), any 
formula for the force would only make sense to the leading order in the separation 
even when the separation is large and subleading terms are meaningless.
It is worth pointing out that a long time ago Gonzales and Estrada-Sarlabons
\cite{gonz1, gonz2} had predicted how the force between such a kink and an 
antikink should decay as a function of the separation between them. However, 
they did not have similar prediction for the kink-kink force. Further, they 
had no specific prediction for the numerical coefficient appearing in the formula. 
In a remarkable paper, Manton \cite{Manton19} gave a detailed prescription for
the calculation of the force between the two well separated kinks and a
well separated kink and antikink pair in the case of the potential (discussed in 
Sec. III)
\be\label{10.1}
V(\phi) = \frac{1}{2} \phi^4 (1-\phi^2)^2\,.
\ee
He not only predicted that both the forces should vary as the inverse fourth 
power of the distance between the two well separated kinks (or the kink and 
antikink) but more importantly also calculated the prefactor multiplying this 
exponent in both the cases and showed that remarkably this factor is very 
different in the case of KK and the K-AK forces. 

In a subsequent paper, Christov et al. \cite{cddgkks} generalized this 
calculation and estimated the force between two well separated kinks as well 
as between a kink and an antikink in the one-parameter family of potentials 
(discussed in Sec. III)
\be\label{10.2}
V(\phi) = \frac{1}{2} \phi^{2n+2} (1-\phi^2)^2\,,~~n = 1, 2, 3,... \,.
\ee
They showed that both the KK and K-AK forces decay like $x^{-2(n+1)/n}$ 
where $x$ is the distance between the two kinks or between the kink and 
the antikink thereby 
reconfirming the prediction of Gonzales and Estrada-Sarlabons [4,5]. 
They also estimated the prefactor multiplying the exponent and 
confirmed that this exponent is indeed very different in the case of the KK 
and K-AK forces. Further, they
compared their predictions with a detailed numerical computation in the specific 
cases of $n = 1, 2, 3$. Note that $n = 1$ is the case studied by Manton
while $n = 2, 3$ correspond to the $\phi^{10}$ and $\phi^{12}$ models, respectively. 
 
\subsection{KK and K-AK Forces For the Model Given By Eq. (\ref{10.1})}

In this review article we will only briefly discuss the key points of the 
$\phi^8$ case, the details can be found in \cite{Manton19, cddgkks} as well as 
in \cite{christov, peru}. In his paper, Manton \cite{Manton19} has discussed 
two different approaches for estimating the KK and K-AK forces and showed 
that to the leading order, both approaches give a similar answer for the two forces. 
In the first approach he calculates the force exerted on one kink by using a version 
of the Noether's theorem to calculate the rate of change of its momentum 
\cite{Manton19}. This is equivalent to using the
energy-momentum tensor (introduced in Sec. II, see Eqs. (\ref{2.2a}) and  
(\ref{2.2b})) to estimate the stress exerted on the half line containing the
kink. In the second approach he tries to approximately solve the full
time-dependent field equations. A simpler but cruder approximation 
\cite{gonz1, gonz2} is to just set up a static field configuration that
incorporates both kinks, satisfying the appropriate boundary conditions.
The nontrivial part in both methods is about the ansatz for the 
interpolating field. 

We would like to remind the readers that as discussed in Sec. III, 
the potential 
(\ref{10.1}) admits a kink solution from $\phi = 0$ to $\phi = 1$
(which we denote by $\phi_{0,1}(x)$), a mirror kink from $-1$ to $0$ (which we
denote by $\phi_{-1,0}(x)$), an antikink from $1$ to $0$ (which we denote by 
$\phi_{1,0}(x)$) and a mirror antikink from $0$ to $-1$ (which we denote by 
$\phi_{0,-1}(x)$). Note that while the kink tail around $\phi = 0$ has a fall off 
like $x^{-1}$, the kink tail around $\phi = \pm 1$ has an exponential fall off. 
Manton in his paper \cite{Manton19} has calculated the force between the mirror 
kink $\phi_{-1,0}(x)$ and the kink $\phi_{0,1}(x)$. He has also calculated
the force between the antikink $\phi_{1,0}(x)$ and the kink $\phi_{0,1}(x)$. 
As shown in Sec. III, the kink energy (i.e. mass) in this case is $2/15$ (see
Eq. (\ref{3.9})). Further, the implicit kink solution $\phi_{0,1}(x)$ as given 
by Eq. (\ref{3.4}) is
\be\label{10.3}
x-A = -\frac{1}{\phi}+ \frac{1}{2}\ln\frac{1+\phi}{1-\phi}\,,
\ee
where $A$ can be thought of as the position of the kink. From here it is 
straightforward to obtain the asymptotic behavior
\bea\label{10.4}
\lim_{x \rightarrow -\infty} \phi(x) \simeq \frac{1}{A-x} +O(\frac{1}{(A-x)^3})\,,
~~~ 
\lim_{x \rightarrow \infty} \phi(x) \simeq 1 -2 e^{-2[x+1-A-(1/2)\ln(2)]}\,.
\eea
Now if $\phi(x-A)$ is the kink solution then the mirror kink solution can be
shown to be $-\phi(-x-A)$. This is because both the kink and the mirror kink
obey the same Bogomolnyi Eq. (\ref{3.4}). 

Let us assume that the kink is located at $A$ and mirror kink at $-A$ with
$A \gg 0$. Let us now split the spatial line at $-X$ and $X$ with $0 \ll x \ll A$
so that for $x < -X$ we have an exact mirror kink field and for $x > X$ we have
an exact kink field. For the intermediate region $-X < x < X$ one assumes that
the interpolating field has a linear behavior i.e. $\phi(x) = \mu x$ which leads 
to $X = A/2$ and $\mu = 4/A^2$. It is then straightforward to calculate the 
kink energy to the leading order in $1/A$ and one
finds the repulsive force between the kink and the mirror kink to be (note
the force is the negative derivative of the energy with respect to the
separation of the kink and the mirror kink) \cite{Manton19}
\be\label{10.5}
F_{KK} = \frac{32}{5 A^4}\,.
\ee
This calculation is conceptually easy to follow and gives expected dependence of
$F$ on $A$ although the coefficient $32/5$ is not so accurate. Proceeding in the
same way, and assuming that a well separated kink is located at $A$ and an 
antikink at $-A$, Manton goes on to calculate the attractive force between the kink 
and the antikink. One assumes that the field is symmetric in $x$ at all times. Further, 
he splits up the spatial line at +$A/2$ and $-A/2$ so that for 
$x \le -A/2$ one has an exact antikink field while for $x \ge A/2$
one has an exact kink field. In between $-A/2 \le x \le A/2$, the
interpolating field is assumed to have quadratic behhaviour, i.e.
$\phi(x) = \alpha + \beta x^2$ where $\alpha$ and $\beta$ are 
determined by demanding that $\phi(x)$ is continuous and has continuous
first derivative at $x = \pm A/2$. One finds $\alpha = 1/A$, $\beta = 4/A^3$.
It is then straightforward to calculate the kink-antikink energy to the
leading order in $1/A$ and the attractive force between the kink and the
antikink turns out to be \cite{Manton19} 
\be\label{10.6}
F_{K-AK} = -\frac{88}{105 A^4}\,.
\ee

Manton then goes on to use an alternative approach \cite{Manton19} where he 
models the kink by a field of the form
\be\label{10.7}
\phi(x,t) = \xi(y)\,,~~y = x-A(t)\,.
\ee
The acceleration $a$ then is $a = \ddot{A}$ which is 
assumed to be small.
Further, one also assumes that squared velocity $(\dot{A})^{2}$ is small 
compared to $1$ so that the motion is nonrelativistic and the Lorentz 
contraction as well as the radiation can be neglected. On substituting the 
accelerating field ansatz in the field Eq. (\ref{2.2}) we get
\be\label{10.8}
\xi''(y)+a\xi'(y) -\frac{dV(\xi)}{d\xi} = 0\,.
\ee
Thus the profile $\xi(y)$ satisfies a static equation which depends on the
acceleration $a$ and evolves adiabatically with time as $a$ varies.
Manton assumes that $a$ varies slowly with 
time and the effect of the term $a\xi'(y)$ is to change the effective potential
to $\hat{V}(\xi) = V(\xi) +\frac{2a}{15}$ where the factor $2/15$ is just  the
kink mass (see Eq. (\ref{3.9})) so that on integrating Eq. (\ref{10.8}) once, 
one obtains in the long range region a simplified equation 
\be\label{10.9}
\frac{d\xi(y)}{dy} = \sqrt{\xi^4+\frac{4a}{15}}\,.
\ee
On integrating it from $\xi = 0$ to $\xi = \infty$ this yields the 
acceleration \cite{Manton19}
\be\label{10.10}
a = \ddot{A} = \frac{44}{A^4}\,,
\ee
and hence the KK force is
\be\label{10.11}
F_{KK} = \frac{2}{15} a = \frac{5.91}{A^4}\,.
\ee
Note that this number is different from the estimate obtained in Eq. 
(\ref{10.5}) by using the static field approach. Manton has given justification 
as to why this number is more reliable than that given by Eq. (\ref{10.6}).

By using a similar approach Manton has also calculated the attractive 
interaction between the $\phi_{1,0}(x)$ antikink and the $\phi_{0,1}$ kink.
The only difference is that since the kink and the antikink attract, hence unlike
the KK case, now $a = -\ddot{A}$. While most of the steps are similar to the
KK force calculation, one crucial difference is that now instead of Eq. 
(\ref{10.9}) one gets the equation
\be\label{10.12}
\frac{d\xi(y)}{dy} = \sqrt{\xi^4-\frac{4a}{15}}\,.
\ee
On integrating $\xi(y)$ from $\xi = (4a/15)^{1/4}$ to $\xi = \infty$ one then 
obtains the acceleration and hence the corresponding K-AK force \cite{Manton19}
\be\label{10.13}
F_{K-AK} = - \frac{1.48}{A^4}\,.
\ee
One thus finds that unlike the exponential tail case (where the magnitudes of the 
KK and K-AK forces are equal), for the power law kinks of the $\phi^{8}$ model
(\ref{10.1}), the ratio of the magnitude of the K-AK and KK force is about 
$1/4$. This is rather remarkable and one needs to understand why this is so.

\subsection{KK and K-AK Forces for The Family of Potential (\ref{10.2})}

There are several obvious questions. The first question is how good is this 
theoretical prediction? Secondly, can one extend it to the entire one-parameter 
family of potentials given by Eq. (\ref{10.2})? These questions
have been answered by Christov et al. \cite{cddgkks} by extending the Manton 
calculation to the entire family of potentials (\ref{10.2}). They show that for
the KK case one again obtains  Eq. (\ref{10.9}) except one has to replace 
$\xi^{4}(y)$ by $\xi^{2n+2}(y)$
and replace the kink mass $2/15$ by $2/(n+2)(n+4)$ (see Eq. (\ref{3.9})). One 
then finds that the
KK force for the entire family of potentials (\ref{10.2}) is given by
\be\label{10.14}
F_{KK} = \bigg [\frac{[\Gamma(\frac{n}{2(n+1)})] [\Gamma(\frac{1}{2(n+1)}]}{2
\sqrt{\pi} (n+1)} \bigg ] \frac{1}{2A^{(n+1)/n}}\,. 
\ee
The corresponding K-AK force is similarly obtained with the same replacement as
above in Eq. (\ref{10.12}) and one finds \cite{cddgkks}
\be\label{10.15}
F_{K-AK} = -\bigg [\frac{\sqrt{\pi} [\Gamma(\frac{n}{2(n+1)})]}{[\Gamma(
\frac{-1}{2(n+1)}]} \bigg ] \frac{1}{2A^{(n+1)/n}}\,. 
\ee
As expected, for $n = 1$ one retrieves the Manton results for the KK and the  
K-AK force as given by Eqs. (\ref{10.11}) and (\ref{10.13}), respectively. 
Further, on using the well known identities
\be\label{10.16}
\Gamma(a+1) = a \Gamma(a)\,,~~\Gamma(a) \Gamma(1-a) = \frac{\pi}{\sin(\pi a)}\,,
\ee
it is straightforward to show that \cite{cddgkks}
\be\label{10.17}
\frac{F_{K-AK}}{F_{KK}} = - \bigg [\sin\frac{\pi}{2(n+1)}\bigg]^{2(n+1)/n}\,.
\ee
Thus as $n$ increases, the tail becomes progressively longer, i.e. as we go from 
$\phi^{8}$ to $ \phi^{10}$, $\phi^{12}$ and higher order models, one finds that that 
the K-AK force becomes progressively weaker compared to the corresponding KK 
force and whereas this ratio is $-1/4$ for $n=1$, this ratio goes to zero in the limit 
$n\rightarrow\infty$.  
It is worth remembering that for the exponential kink tails, 
the two forces are always equal and opposite. This is a highly nontrivial result 
which needs a deeper understanding.

These theoretical predictions have been compared with the detailed numerical
simulations for the $n = 1, 2, 3$ cases (i.e. $\phi^8$, $\phi^{10}$ and $\phi^{12}$) 
in \cite{cddgkks} (also see \cite{christov}). There are major challenges in even 
initializing kinks with a power law tail numerically. There is a danger that the initial 
conditions in a direct numerical simulation of interactions may substantially 
affect the nature of the observed interactions \cite{gani2}. 
On using the split-domain ansatz and periodic boundary conditions
numerically accurate predictions were obtained for the KK and the K-AK
forces in case $n = 1, 2$ and $3$. They find that 
while the agreement for the exponent is excellent for all the three cases, the 
agreement about the prefactor multiplying the exponent is excellent for the 
$\phi^8$ case (as given by Eqs. (\ref{10.11}) and (\ref{10.13})), but as $n$ 
increases to $2$ and $3$, the agreement about the prefactor
gradually becomes worse. One of the reasons for this disagreement is that as $n$
increases, the kink tail becomes progressively longer. 

Can we extend the above calculations in case the kink tails are of the form
$peep$ or $pppp$ or $ette$ or $pttp$ or $pppe$ or $eeep$? We now show that the 
answer is yes and 
give predictions for the KK and K-AK forces in the above cases when the two 
tails facing each other have a power law fall off.

\subsection{Predictions for $F_{K-AK}$ in the case of $peep$}

Let us consider the one-parameter family of potentials as given by Eq. 
(\ref{4.1}), i.e.
\be\label{10.18}
V(\phi) = \frac{1}{2} \phi^2 (1-\phi^2)^{2n+2}\,,~~n = 1, 2, 3,... \,.
\ee
This model admits a kink $\phi_{0,1}(x)$, a mirror kink $\phi_{-1,0}(x)$
with an exponential tail around $\phi = 0$ and a power law tail around $\phi = 
\pm 1$. It also admits an antikink $\phi_{1,0}(x)$ and a mirror antikink 
$\phi_{0,-1}(x)$. 

Following the procedure discussed above, it is straightforward to calculate 
the force between the kink $\phi_{0,1}(x)$ and the antikink $\phi_{1,0}(x)$.
The two differences compared to the $eppe$ case discussed above are (i) in 
this case in the long range limit the 
potential around $\phi = 1$ is $\frac{1}{2} (2\xi)^{2n+2}$ where 
$\xi = 1-\phi$ and (ii)
the kink mass in this case is $M_k = 1/[2(n+2)]$. Remarkably it turns out that
after taking into account these factors the K-AK force for the $eppe$ case is 
in fact identical to the K-AK force in the $eppe$ case as discussed above 
provided we replace $A$ by $2A$ in Eq. (\ref{10.15}) and is given by
\be\label{10.15a}
F_{K-AK} = -\bigg [\frac{\sqrt{\pi} [\Gamma(\frac{n}{2(n+1)})]}{[\Gamma(
\frac{-1}{2(n+1)}]} \bigg ] \frac{1}{2(2A)^{(n+1)/n}}\,.
\ee
This is rather remarkable and suggests that the KK and K-AK forces are 
independent of the kink mass and only depend on the asymptotic behavior of 
the power law kink tail. Let us now check if the same is borne out in case the 
tails are of the form $pppp$. 

\subsection{Predictions for $F_{KK}$ and $F_{K-AK}$ in the case of $pppp$}

Let us consider the one-parameter family of potentials as given by Eq. 
(\ref{5.1}), i.e.
\be\label{10.19}
V(\phi) = \frac{1}{2} \phi^{2m+2} (1-\phi^2)^{2n+2}\,,~~m,n = 1, 2, 3,... \,.
\ee
This model admits a kink $\phi_{0,1}(x)$, a mirror kink $\phi_{-1,0}(x)$
with a power law tail around $\phi = 0$ as well as around $\phi = 
\pm 1$. It also admits an antikink $\phi_{1,0}(x)$ and a mirror antikink 
$\phi_{0,-1}(x)$. In particular, while the tail around $\phi = 0$ 
asymptotically goes like $x^{-1/m}$, the power law tails around $\phi = \pm 1$
go like $x^{-1/n}$. 

Following the procedure discussed above, it is straightforward to calculate 
the force between the mirror kink $\phi_{-1,0}(x)$ and the kink $\phi_{0,1}(x)$
as well as the force between the antikink $\phi_{1,0}(x)$ and the kink 
$\phi_{0,1}(x)$ by noting that in this case the potential around $\phi = 0$
is given by $\frac{1}{2} \phi^{2m+2}$ while the kink mass in this case is as 
given by Eq. (\ref{5.14}). Remarkably it  turns out that after taking into account 
these factors the K-K and K-AK forces and 
their ratio are again given by the same expressions as in the $eppe$ case 
provided we make the obvious replacement of  $n$ by $m$ in Eqs. (\ref{10.14}), 
(\ref{10.15}) and (\ref{10.17}).

In this model one can also calculate the force between the kink $\phi_{0,1}(x)$
and the antikink $\phi_{1,0}(x)$ since the kink tail around $\phi = 1$ also has a
power law tail. On noting that in the long range limit the potential around 
$\phi = 0$ is given by $\frac{1}{2} (2\xi)^{2n+2}$ where $\xi = 1-\phi$, it is
straightforward to check that the K-AK force in this case is 
also the same as in the $eppe$ case provided we replace $A$ by $2A$ in 
Eq. (\ref{10.15}) and is given by Eq. (\ref{10.15a}).

Finally we come to the question of the KK and K-AK forces in the model with 
power-tower kink tails as discussed in 
Sec. VI. As has been discussed in detail in \cite{KS20}, the power-tower tails
essentially are power law tails varying like $x^{-1/a}$ with $a$ 
being a real positive number. We would like to remark that the calculation of 
 \cite{cddgkks} which is valid when the kink tail goes like $x^{-1/n}$ where $n$ is
an integer, is easily extended to the case where the tail falls off like $x^{-1/a}$ 
 and $a$ is a real positive number. In particular, the expressions for the KK 
 force, the K-AK force and their ratio continue to be still given by Eqs. 
 (\ref{10.14}), (\ref{10.15}) and (\ref{10.17}) with the obvious replacement 
 of $n$ by $a$. As has been shown in \cite{KS20}, for potentials
 \be\label{10.20}
 V(\phi) = \frac{1}{2} \phi^4 \bigg[\frac{1}{2} \ln(\phi^2)\bigg]^{2m}\,,
 \ee
 where the kink tails are of the form $ette$, the kink tail around $\phi = 0$ 
 falls off like $x^{-1/a}$ where $0 < a < 1$ and as $m$ increases the value of $a$ 
 decreases progressively. It appears that these potentials may 
 provide a bridge between the kink solutions with a power law tail and the kink 
 solutions with an exponential tail. It would be highly desirable to understand this 
 transition. 
 
Proceeding in the same way, it is straightforward to compute the KK and K-AK 
forces and their ratio in the case of the other models discussed in this review.

\section{Kink-Antikink Collisions at Finite velocity}

The issue of kink-antikink (K-AK) collisions at finite velocity is only
recently being addressed in the context of the kinks with a power law tail
\cite{gani2,gani21,chr21,camp}. Before we discuss the results obtained, 
nontrivial issues involved and some of the open problems, it is worth
pointing out that in the context of kinks with an exponential tail, the
K-AK collisions at finite velocity have been extensively discussed 
during the last four and half 
decades \cite{gani14,kud,sug,cam,goo,bel,gani99,bazx,nie,moh,adam,zhong,yan,
man21,per,liz}. However, it is fair to say that a coherent physical explanation is 
still lacking. Some of the key findings are the following. 

\begin{enumerate}

\item There is a critical value $v_{cr}$ of the initial velocity which is
model dependent but typically of the order of $0.1-0.3$ (in units of $c$, the
velocity of light) separating two different regimes of collision: for $v_{in}
< v_{cr}$, the capture and the formation of a bound state occurs, while for
$v_{in} > v_{cr}$, the kink and the antikink escape to infinity after a single
collision. 

\item For $v_{in} < v_{cr}$, besides the formation of a bound state one finds 
that there are two-bounce, three-bounce and so on escape windows. In 
particular, there are intervals of the initial velocities within which kinks 
scatter and eventually escape to spatial infinities. The difference from 
$v_{in} > v_{cr}$, however, is that inside the escape windows the kinks scatter 
to infinity not after a single impact, but even after two or more successive
collisions. The escape windows seem to form a fractal structure.

\item The explanation of the escape windows phenomenon seems to vary from model 
to model. For those models (like $\phi^4$) where the kink stability equation 
admits an extra mode called the vibrational mode (i.e. over and above the 
translational zero mode which is present in every kink bearing model), it has 
been suggested that the above phenomenon occurs due to the resonant energy 
exchange between the zero mode and the vibrational mode \cite{liz}. 

\item On the other hand, models where the kink stability equation does not 
have any vibrational mode (like the celebrated $\phi^6$ model with 
	$V(\phi) = \frac{1}{2} \phi^2 (1-\phi^2)^2$), it has been 
proposed that the escape windows occur because the kinks and the antikinks 
of this model are asymmetric \cite{dor}.  
\end{enumerate}

During the last few years researchers have inquired \cite{gani2,gani21,chr21,camp} if 
similar behavior also occurs in the collision of power law kinks at finite 
velocity. The major challenge in case the kinks facing each other have a power 
law tail is the correct formulation of the initial conditions since the power 
law kink tails overlap significantly at any finite distance that can be considered 
as asymptotically large, i.e. large enough so that the kink and the antikink could 
be considered noninteracting. Using the split-domain ansatz \cite{chr21} the
authors have investigated the kink-antikink collisions with finite velocity in
the $\phi^{8}, \phi^{10}$ and $\phi^{12}$ models as given by Eq. (\ref{3.1}) 
with $n = 1, 2, 3$, i.e. 
\be\label{11.1}
V(\phi) = \frac{1}{2} \phi^{4,6,8} (1-\phi^2)^2\,.
\ee
In these models there is a kink from $0$ to $1$, a mirror kink from $-1$ to $0$
and the corresponding two antikinks with a power law kink tail around $\phi = 0$ 
and an exponential tail around $\phi = \pm 1$. These authors have considered 
collisions between the kink $\phi_{k}(-1,0)$ and the antikink $\phi_{k}(0,-1)$ moving 
at finite velocity. In all these models they again find the same qualitative behavior 
as in the collision of the kink and the antikink with an exponential tail. In particular, 
they again find that there is a critical velocity $v_{cr}$ such that for $v_{in}
> v_{cr}$ the kinks escape to infinity after one impact while for $v_{in} <
v_{cr}$ there is formation of a kink-antikink bound state. Besides that, they 
again find two-bounce, three-bounce and so on escape windows as depicted 
in Fig. 9. The critical velocity $v_{cr}$ monotonically increases from $v_{cr}$ of 
about $0.15$ for $\phi^{8}$ case to about $0.21$ for $\phi^{10}$ and to $0.27$ for 
the $\phi^{12}$ case. 

\begin{figure}[h] 
\includegraphics[width=6.0 in]{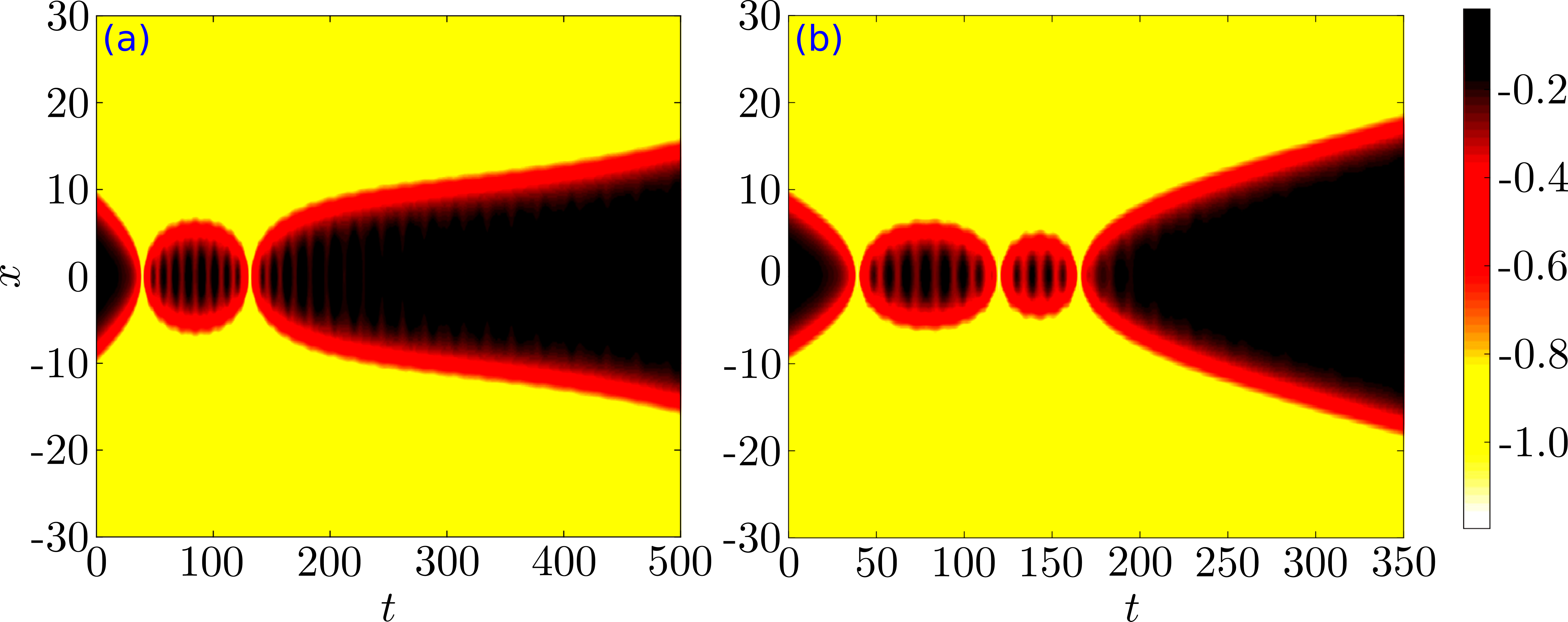}
\caption{Space-time contour plot of the scalar field of the kink-antikink interaction   
in the $\phi^8$ model showing (a) two-bounce and (b) three-bounce windows. 
Adapted from \cite{chr21}. }
\end{figure} 

For the power law kinks, this area of research is in its infancy and hopefully 
in coming years we will have a better understanding and eventually a coherent 
physical explanation might be available. Some of the open problems are the 
following.

\begin{enumerate}

\item For the models as given by Eq. (\ref{3.1}), 
for large $n$ do we still have a similar picture as for $n = 1, 2, 3$? In
particular as $n$ becomes progressively larger, does $v_{cr}$ become 
progressively larger and approach~1 (in units of $c$, the velocity of light) 
asymptotically? For large values of $v_{cr}$, how important are the relativistic 
effects? Of course this a difficult problem since as $n$ increases the kink 
tails become progressively longer.

\item The question of $v_{cr}$ being very large is not a hypothetical question 
because of a recent study \cite{camp} of K-AK collisions at finite velocity 
for different models $\phi^{8,12}$ as given by Eq. (\ref{5.15}) 
with $n = 1, 2$ (see Fig. 10), i.e.   
\be\label{11.3}
V(\phi) = \frac{1}{2} (1-\phi^2)^{4,6}\,.
\ee
Note that in these models there is only a single kink from $-1$ to +$1$ and a 
corresponding antikink and unlike the model studied by \cite{chr21}, in these
models one has a symmetric power law kink tail around $\phi = 1$ and $\phi = -1$.
Using what they term as computationally efficient way, it has been claimed by 
the authors of \cite{camp} 
that the K-AK collisions at finite velocity behave very differently from all 
the models studied previously (be with an exponential or a power law tail). In 
particular, they claim that their is neither a long-lived bound state formation
nor resonance windows and $v_{cr}$ is ultrarelativistic. These claims, if true, 
would indicate that for power law kink tails, the K-AK collisions at finite 
velocity are conceptually very different from the exponential tail case. 

\begin{figure}[h] 
\includegraphics[width=5.8 in]{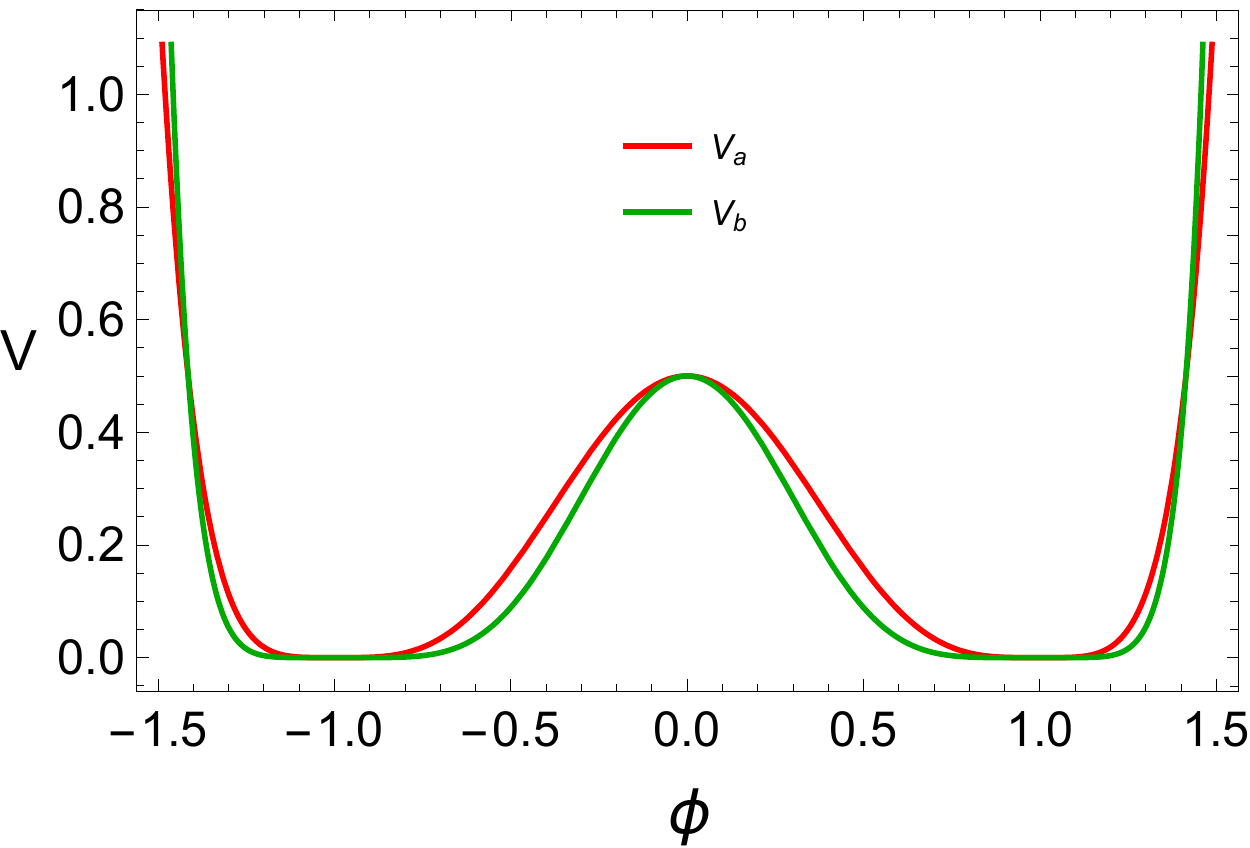}
\caption{The $\phi^8$ ($V_a$) and $\phi^{12}$ ($V_b$) potentials given by Eq. (146). }
\end{figure} 

\item Apart from the two possible kind of models studied so far \cite{chr21,
camp}, there are several other models with power law kink tails as discussed
in sections IV to IX and one also needs to study K-AK collisions at finite velocity
in these models and try to understand various facets of this challenging
as well as difficult problem.

\end{enumerate}
Finally, we note that recently kink scattering in a generalized Wess-Zumino 
model \cite{WZ} with three minima has been studied, wherein two different 
scattering channels have been identified, namely kink-kink reflection and 
kink-kink hybridization. 

\section{Open Problems}

In this review we have tried to uncover various aspects of kink solutions with 
a power law tail. Several issues are now fairly clear. For example, one distinctive 
feature of kinks with a power law tail is that the corresponding kink stability 
equation only admits the translational zero mode and
in these cases the beginning of the continuum coincides with the zero mode 
(i.e. with $\omega^2 = 0$). Secondly, there is a recipe available for constructing
two adjoining kink solutions with various possible combinations of the 
power law and the exponential kink tails. Besides, a wide class of kink solutions 
has been constructed with two of the tails being of power-tower type \cite{KS20} 
which is a kind of power law tail behaving like $x^{-1/a}$ where $a$ is any arbitrary
positive number. However, there are several aspects of the kink solutions with
the power law tail which are either only partially understood or not understood
at all. We now list some of these issues.

\begin{enumerate}

\item So far only implicit kink solutions with a power law tail have been 
constructed in models with polynomial potentials. While explicit kink 
solutions with a power law tail have been constructed in a few models with 
polynomial potentials, but unfortunately in all these models, while the
kink potential $V(\phi)$ is continuous but its derivative is not. Even 
though that has still allowed the explicit construction of the kink solutions
with a power law tail, it is certainly desirable to obtain kink solutions with
a power law tail in models with polynomial potentials where not only the 
kink potential but its derivative is also continuous everywhere. 

\item One of the major issues which is only partially understood is that of the 
KK and K-AK forces in the 
case of the power law tail. In case the kink tails are of the form $eppe$, one
of the intriguing conclusions \cite{Manton19, cddgkks} is that compared to the 
kink-kink force, the kink-antikink force gets progressively weaker as the kink 
tail becomes progressively longer. This is in contrast to the kinks with 
exponential tails for which the magnitudes of the KK and K-AK forces are 
always equal. In our view, understanding the reason for the weak K-AK force 
compared to the KK force remains one of the major open problems.

\item The calculation of \cite{cddgkks} for the KK and K-AK forces is only
reliable for smaller values of $n$ in the family of potentials given by 
Eq. (\ref{10.2}) since as $n$ increases, the kink tails become progressively 
longer. It is highly desirable to devise methods for a reliable computation of 
the KK and K-AK forces for the entire family of potentials as given by
Eq. (\ref{10.2}).

\item Are the results derived by \cite{Manton19, cddgkks} for the $eppe$ case
for the KK and K-AK forces and their ratio also valid for the other kink tail 
configurations such as $peep$, $pppp$, etc. (with suitable modification for the 
form of the potential in the asymptotic limit) and the corresponding kink mass? 
In Sec. X we have assumed this to be true and tried to make predictions for the 
KK and K-AK forces and their ratio in the other cases with power law kink tails. 
How good are these predictions? It would be desirable if these could be 
checked by accurate numerical calculations. 
		
\item The predictions for the KK and K-AK forces have been made in \cite{Manton19,
cddgkks} for the $eppe$ case, which indicated that the kink tail goes like $x^{-1/n}$ 
where $n$ is any integer. Are these results also valid in case the kink tail goes like 
$x^{-1/a}$ with $a$ being any positive number? In Sec. X we have assumed it to 
be true and tried to make predictions for the KK and K-AK forces in case the kink 
tails are of the form $ette$ or $pttp$ and on the basis of these predictions have 
suggested that the power-tower kink tails \cite{KS20} form a bridge between the 
exponential and the power law tails. How good are these predictions? It would be 
desirable if these could also be checked by accurate numerical calculations. 

\item The predictions made in \cite{Manton19, cddgkks} for the KK and K-AK 
forces are only valid to the leading order and only for the large separation 
between the two kinks or the kink and the antikink. Is it possible to compute the 
subleading corrections to these estimates? 

\item In all the calculations an implicit assumption has been made that these 
results do not depend on the effect of the two (of the four) tails which are
not facing each other. While it may be a reasonable assumption in the $eppe$ 
case where the other two tails are exponential, it is not obvious if it is
also true in case the kink tails are of the form $pppp$. It would be highly 
desirable if one can numerically check the validity of this prediction.

\item Normally for exponential tails, the KK and K-AK forces fall off  
exponentially as a function of the distance between them. Is this 
conclusion still valid if the two remaining kink tails have a power law fall off 
as in the $peep$ case?

\item Can one extend the calculation of Manton \cite{Manton19} to the case
of the nonpolynomial models \cite{bmm, mra, ko} with a power law kink tail?
Knowing the behaviour of the power law kink tail, one can perhaps predict
how the K-K and K-AK forces fall off as a function of the kink-kink
(or kink-antikink) separation. However, it is not at all obvious how to compute 
the pre-factor multiplying it using the Manton approach \cite{Manton19}.

\item There are several areas of physics where kink solutions with a power law tail 
could have applications. However, to date we do not have any such concrete examples. 
Finding a concrete example will give an added motivation for an in-depth study of  
kinks with power law tails.

\item Very little is known about the kink-antikink collisions at finite 
velocity in the case of kinks with power law tails. In fact, one
has conflicting results about such collisions for the $eppe$ case \cite{chr21} 
compared to the case where there is a single kink with symmetric power law 
kink tails at both the ends \cite{camp}. It is clearly important to study such 
collisions in other models discussed in sections IV to IX where too one has kink 
solutions with a power law tail. 

\item Recently we have also constructed models having super-exponential (se) 
\cite{PKS} and super-super exponential (sse) \cite{KS20a} kink tails. It may be 
of interest to enquire if one can construct models where the two adjoining kinks
have an arbitrary combination of the power law, the exponential, the 
super-exponential and the super-super-exponential tails.

\item Several years ago Bazeia et al. \cite{baz02} proposed a novel 
deformation function $f(\phi) = \sqrt{1-\phi^2}$ and discussed some of its
properties. Subsequently, a lot of work has been done about various other 
deformation functions \cite{baz06, alm, baz061,bli, deb, baz18}. Recently 
we \cite{KS21a} have generalized the deformation function of \cite{baz02} 
and proposed a one-parameter family of deformation functions 
$f(\phi) = (1-\phi^{2n})^{1/2n}, n = 1, 2, 3,... $ having novel and very 
unusual properties such as being its own inverse and starting from certain
potentials such a deformation can either create or destroy an arbitrary even 
number of kink solutions. It would be worthwhile exploring the connection 
between the various models with power law kink tails such as those mentioned 
in sections III to IX using this and other deformation functions.

\end{enumerate}

We hope that in the coming years insightful answers would be obtained for at
least some of the questions raised above, further enriching the field of power 
law kink tails.

{\bf Acknowledgments.} We thank Ayhan Duzgun for help with the figures. One 
of us AK is grateful to Indian National Science Academy (INSA) for the award of 
the INSA Senior Scientist position at Savitribai Phule Pune University, Pune, India. 
The work at Los Alamos National Laboratory was carried out under the auspices 
of the U.S. DOE and NNSA under Contract No. DEAC52-06NA25396.

\end{document}